\documentclass[journal=jacsat,manuscript=article]{achemso}
\usepackage{amsmath,amssymb}
\usepackage[version=4]{mhchem}
\usepackage{hyperref}
\usepackage{xcolor}
\usepackage{gensymb}

\author{Maryam Ali}
\affiliation{Institute of Physical Chemistry, Friedrich-Schiller-Universit\"at Jena, 07743 Jena, Germany}
\alsoaffiliation{Leibniz Institute of Photonic Technology, Albert-Einstein-Straße 9, 07745 Jena, Germany}
\author{Robin Schneider}
\affiliation{Leibniz Institute of Photonic Technology, Albert-Einstein-Straße 9, 07745 Jena, Germany}
\alsoaffiliation{Aalen University of Applied Sciences, Beethovenstraße 1, 73430 Aalen, Germany}
\author{Anika Strecker}
\affiliation{Leibniz Institute of Photonic Technology, Albert-Einstein-Straße 9, 07745 Jena, Germany}
\alsoaffiliation{Ernst-Abbe University of Applied Sciences, 07745 Jena, Germany}
\author{Nila Krishnakumar}
\affiliation{Abbe Center of Photonics, Friedrich-Schiller-Universit\"at Jena, 07745 Jena, Germany}
\alsoaffiliation{Leibniz Institute of Photonic Technology, Albert-Einstein-Straße 9, 07745 Jena, Germany}
\author{Sebastian Unger}
\affiliation{Institute of Physical Chemistry, Friedrich-Schiller-Universit\"at Jena, 07743 Jena Germany}
\alsoaffiliation{Leibniz Institute of Photonic Technology, Albert-Einstein-Straße 9, 07745 Jena, Germany}
\author{Mohammad Soltaninezhad}
\affiliation{Institute of Physical Chemistry, Friedrich-Schiller-Universit\"at Jena, 07743 Jena, Germany}
\alsoaffiliation{Leibniz Institute of Photonic Technology, Albert-Einstein-Straße 9, 07745 Jena, Germany}
\author{Johanna Kirchhoff}
\affiliation{Institute of Physical Chemistry, Friedrich-Schiller-Universit\"at Jena, 07743 Jena, Germany}
\alsoaffiliation{Leibniz Institute of Photonic Technology, Albert-Einstein-Straße 9, 07745 Jena, Germany}
\alsoaffiliation{Jena University Hospital, 07747 Jena, Germany}
\author{Astrid Tannert}
\affiliation{Jena Biophotonics and Imaging Laboratory, 07745 Jena, Germany}
\alsoaffiliation{Leibniz Institute of Photonic Technology, Albert-Einstein-Straße 9, 07745 Jena, Germany}
\alsoaffiliation{Jena University Hospital, 07747 Jena, Germany}
\author{Katerina A. Dragounova}
\affiliation{Institute of Physical Chemistry, Friedrich-Schiller-Universit\"at Jena, 07743 Jena, Germany}
\alsoaffiliation{Leibniz Institute of Photonic Technology, Albert-Einstein-Straße 9, 07745 Jena, Germany}
\alsoaffiliation{Jena University Hospital, 07747 Jena, Germany}

\author{Rainer Heintzmann}
\affiliation{Institute of Physical Chemistry, Friedrich-Schiller-Universit\"at Jena, 07743 Jena, Germany}
\alsoaffiliation{Leibniz Institute of Photonic Technology, Albert-Einstein-Straße 9, 07745 Jena, Germany}
\alsoaffiliation{Abbe Center of Photonics, Friedrich-Schiller-Universit\"at Jena, 07745 Jena, Germany}
\author{Anne-Dorothea M\"uller}
\affiliation{Anfatec Instruments AG, Melanchtonstraße 28, 08606 Oelsnitz, Germany}
\author{Christoph Krafft}
\affiliation{Institute of Physical Chemistry, Friedrich-Schiller-Universit\"at Jena, 07743 Jena, Germany}
\alsoaffiliation{Leibniz Institute of Photonic Technology, Albert-Einstein-Straße 9, 07745 Jena, Germany}
\author{Ute Neugebauer}
\affiliation{Institute of Physical Chemistry, Friedrich-Schiller-Universit\"at Jena, 07743 Jena, Germany}
\alsoaffiliation{Leibniz Institute of Photonic Technology, Albert-Einstein-Straße 9, 07745 Jena, Germany}
\alsoaffiliation{Abbe Center of Photonics, Friedrich-Schiller-Universit\"at Jena, 07745 Jena, Germany}
\alsoaffiliation{Jena Biophotonics and Imaging Laboratory, 07745 Jena, Germany}
\alsoaffiliation{Jena University Hospital, 07747 Jena, Germany}
\author{Daniela T\"auber}
\email{dantaube@gmx.de}
\affiliation{Institute of Physical Chemistry, Friedrich-Schiller-Universit\"at Jena,  07743 Jena, Germany}
\alsoaffiliation{Leibniz Institute of Photonic Technology, Albert-Einstein-Straße 9, 07745 Jena, Germany}

\title{Nano-chemical cell-surface evaluation in photothermal spectroscopic imaging of antimicrobial interaction in model system \textit{Bacillus subtilis} \& vancomycin}

\begin{document}
\clearpage

\begin{abstract}
The power of photothermal spectroscopic imaging to visualize antimicrobial interaction on the surface of individual bacteria cells has been demonstrated on the model system \textit{Bacillus subtilis} and vancomycin using mid-infrared photo-induced force microscopy (PiF-IR, also mid-IR PiFM). High-resolution PiF contrasts obtained by merging subsequent PiF-IR scans at two different illumination frequencies revealed chemical details of cell wall destruction after 30 and 60~min incubation with vancomycin with a spatial resolution of $\approx 5$~nm. This approach compensates local intensity variations induced by near-field coupling of the illuminating electric field with nanostructured surfaces, which appear in single-frequency contrasts in photothermal imaging methods, as shown by [Anindo \textit{et al., J. Phys. Chem C}, 2025, \textbf{129}, 4517]. Known spectral shifts associated with hydrogen bond formation between vancomycin and the \ce{N-acyl-D-Ala4-D-Ala5} termini in the peptidoglycan cell wall have been observed in chemometrics of PiF-IR spectra from treated and untreated \textit{Bacillus subtilis} harvested after 30~min from the same experiment. Spectral signatures of the vancomyin interaction have been located in the piecrust of a progressing septum with $\approx 10$~nm resolution using PiF contrasts of three selected bands of a PiF-IR hyperspectral scan of an individual \textit{Bacillus subtilis} cell harvested after 30~min incubation. Our results are complemented by a discussion of imaging artifacts and the influence of parameter settings supporting further development towards standardization in the application of PiF-IR for visualizing the chemical interaction of antibiotics on the surface of microbes with few nanometer resolution.
\end{abstract}

\maketitle

\section{Introduction}
The global rise of antimicrobial resistance (AMR) poses an increasing live threat all over the world.\cite{murray_global_2022} Efforts to develop new drugs and therapies will ultimately benefit from the ability to achieve chemical information at the subcellular and single-molecule level. Recently, a number of photothermal spectroscopic imaging methods have been developed, which overcome the spatial limitation of conventional IR spectroscopy and complement insights from electron microscopy (EM), Raman spectroscopic imaging, and fluorescence microscopy on the nanoscale by combining powerful IR illumination with the integration of other detection schemes,\cite{xiao_spectroscopic_2018, wang_toward_2023, wang_super-resolution_2022, mathurin_photothermal_2022, schwartz_guide_2022, jahng_substructure_2018, davies-jones_photoinduced_2023, joseph_nanoscale_2024, shcherbakov_photo-induced_2025, dos_santos_nanoscale_2020, kanevche_infrared_2021, shen_scanning_2024} for example, the use of visual probe wavelengths or scanning (atomic) force microscopy (AFM). Among the latter, nano-FTIR and IR scattering scanning nearfield optical microscopy (IR-SNOM) employ optical detection of the near-field absorption in the sample mediated by an AFM tip.\cite{kanevche_infrared_2021, shen_scanning_2024, bakir_orientation_2020, greaves_label-free_2023} In contrast, the so-called AFM-IR techniques combine IR illumination with mechanical detection.\cite{mathurin_photothermal_2022, xiao_spectroscopic_2018, shcherbakov_photo-induced_2025} AFM-IR was initially coined for the oldest of these techniques, photo-thermal induced resonance (PTIR). Among the various AFM-IR techniques, PTIR,\cite{mathurin_photothermal_2022, schwartz_guide_2022, kochan_atomic_2020, banas_comparing_2021} mid-IR peak force microscopy (PFIR)\cite{wang_super-resolution_2022, wang_principle_2022, xie_dual-frequency_2022} and tapping AFM-IR\cite{mathurin_photothermal_2022, schwartz_guide_2022, hondl_method_2025} probe the thermal expansion upon IR absorption in the sample, whereas, mid-IR photo-induced force microscopy (PiF-IR or mid-IR-PiFM)\cite{shcherbakov_photo-induced_2025, davies-jones_photo_2022, joseph_nanoscale_2024, ji_label-free_2019, jahng_substructure_2018, sifat_photo-induced_2022, anindo_photothermal_2025} probes the induced change in the electromagnetic near-field.\cite{jahng_nanoscale_2019, jahng_quantitative_2022, sifat_photo-induced_2022, anindo_photothermal_2025}
 Having been developed for applications in the Material Sciences, meanwhile, an increasing number of applications of photothermal imaging in the Life Sciences has been reported.\cite{wang_toward_2023, wang_super-resolution_2022, mathurin_photothermal_2022, davies-jones_photoinduced_2023, dos_santos_nanoscale_2020, kanevche_infrared_2021, shen_scanning_2024, bakir_orientation_2020, greaves_label-free_2023, kochan_atomic_2020, banas_comparing_2021, davies-jones_photo_2022, joseph_nanoscale_2024, shcherbakov_photo-induced_2025, ji_label-free_2019, abrego-martinez_aptamer-based_2022, zancajo_ftir_2020, kochan_detection_2019,hondl_method_2025} Among those methods, PiF-IR stands out by its unprecedented spatial resolution of $\approx 5$ nm, which is combined with a high spectral resolution of 1 ${\rm cm}^{-1}$.\cite{davies-jones_photoinduced_2023, davies-jones_photo_2022, joseph_nanoscale_2024, krishnakumar_nanoscale_2019, jahng_substructure_2018, sifat_photo-induced_2022, murdick_photoinduced_2017, shcherbakov_photo-induced_2025} The exceptional lateral spatial resolution and surface-sensitivity of PiF-IR is achieved by combining non-contact AFM mode with electronic filtering using a heterodyne detection scheme of the force gradient between a sharp metallic AFM tip and the sample.\cite{davies-jones_photoinduced_2023, shen_scanning_2024, sifat_photo-induced_2022, anindo_photothermal_2025} In a recent study, we applied PiF-IR to polymerized Actin, a major structural protein in eucaryotic cells, demonstrating $\approx 5$ nm spatial and submolecular spectral resolution in hyperspectral PiF-IR images.\cite{joseph_nanoscale_2024}

Several studies have been published applying photothermal imaging methods to microbe investigation. Kochan \textit{et al.}\ used PTIR to study single cells of Gram-positive \textit{Staphylococcus aureus} (\textit{S.\ aureus}) demonstrating subcellular details with 20-100 nm resolution.\cite{kochan_atomic_2020} 
The cell wall of Gram-positive bacteria consists of a thick layer of the polymer peptidoglycan which is made up of glycan strands that are cross-linked by peptide side chains,\cite{hayhurst_cell_2008} however, slight differences in the organization of the glycan strands between different species of Gram-positive bacteria have been reported in studies using electron microcopy\cite{tulum_peptidoglycan_2019, pajerski_attachment_2019} or high-resolution AFM.\cite{touhami_atomic_2004,hayhurst_cell_2008} Applying quick-freeze, deep-etch EM, Tulum \textit{et al.}\ found thinner filaments (width $\approx 7$ nm) arranged concentrically around the poles, while thicker filaments (width $\approx 9$ nm) were aligned in a partially circumferential manner on the cylindrical part of the \textit{B.~subtilis} cells\cite{tulum_peptidoglycan_2019} in agreement with other studies on \textit{B.~subtilis}.\cite{hayhurst_cell_2008} In contrast, in the cell wall of \textit{S.\ aureus}, peptidoglycan strands have been observed to form a loosely arranged fiber network with a large number of empty spaces between them. \cite{touhami_atomic_2004,filip_ft-ir_2004} In their study, Kochan \textit{et al.}\ demonstrated the chemical variation between a PTIR spectrum obtained on a septum preceding cell division of a \textit{S.\ aureus} cell and another PTIR spectrum recorded from cell area, whereas, individual glycan strands were not resolved.\cite{kochan_atomic_2020}
In an earlier study, Kochan \textit{et al.}\ had applied PTIR to characterize molecular changes in strains of \textit{S.\ aureus} associated with resistance towards either vancomycin, a glycopeptide antibiotic, or daptomycin, a lipopeptide antibiotic.\cite{kochan_detection_2019} Comparing parent to resistant strains of paired clinical isolates on a single-cell level combined with chemometrics in data analysis, they reported an increase in the amount of intracellular carbohydrates for the strain showing intermediate resistance to vancomycin. An increase in intensity was found in the spectral region between 1200-1000~${\rm cm}^{-1}$ (phosphodiester stretching vibrations and several carbohydrate modes, both present in glycans) relative to the amide I band (1600-1690~${\rm cm}^{-1}$) for resistant strains compared to parent strains (susceptible), which the authors interpreted as thicker peptidoglycan layers of resistant strains.\cite{kochan_detection_2019} Cells of the strain showing resistance to daptomycin additionally showed an increase in the lipid content.\cite{kochan_detection_2019} These molecular changes could not be observed from investigation of the same strains using attenuated total reflectance (ATR) infrared spectroscopy, which is not sensitive to the investigation of single bacteria cells, but requires larger sample volumes.\cite{kochan_detection_2019} Davies-Jones \textit{et al.}\ applied PiF-IR to single cells of \textit{E. coli}, \textit{S.\ aureus} and the yeast \textit{Candida albicans} revealing nanoscale chemical contrasts of cell walls and across sectioned cells together with a considerably higher spectral resolution compared to that of obtained FTIR spectra of the three microbes.\cite{davies-jones_photoinduced_2023} Hondl \textit{et al.}\ used tapping AFM-IR to study extracellular vesicles (EVs) extracted from human milk providing chemical contrasts at a spatial resolution of $\approx 20$~nm.\cite{hondl_method_2025} EVs play a key role in intercellular communication between various cell types, including also microbes such as \textit{Bacillus subtilis} (\textit{B.~subtilis}).\cite{brown_extracellular_2014, kim_extracellular_2016, toyofuku_prophage-triggered_2017}

However, in several studies using various photothermal imaging methods, anisotropic intensity distributions have been reported from the investigation of nanostructured surfaces, including the study of EVs by Hondl \textit{et al.}\cite{anindo_photothermal_2025, hondl_method_2025, xie_dual-frequency_2022} In a recent study, Anindo \textit{et al.}\ combined theoretical modeling with experimental investigation using PiF-IR to demonstrate coupling effects of the illuminating electric field with nanostructured dielectric materials absorbing at mid-IR frequencies\cite{anindo_photothermal_2025} extending an earlier work by Xie \textit{et al.}\cite{xie_dual-frequency_2022} Such near-field coupling results in spatial anisotropies in photothermal imaging.\cite{anindo_photothermal_2025, hondl_method_2025, xie_dual-frequency_2022} Anindo \textit{et al.}\ investigated the appearance of these effects for illumination frequencies within a broad spectral range covering several absorption bands of the sample material as well as non-absorbing spectral regions. At sufficiently low illumination power, the observed spatial anisotropies did not show an effect on the spectral shapes of the PiF-IR spectra acquired from a PMMA nanosphere with 100~nm diameter allowing for a qualitative interpretation of the photothermal images.\cite{anindo_photothermal_2025}

In this work, we present the application of PiF-IR to a well-known model system for antibiotic interaction: Gram-positive \textit{B. subtilis},\cite{hayhurst_cell_2008, tulum_peptidoglycan_2019, filip_ft-ir_2004, stein_bacillus_2005} treated with vancomycin.\cite{kahne_glycopeptide_2005} Vancomycin is a $\beta$-lactam antibiotic that affects cell wall growth\cite{kahne_glycopeptide_2005, wong_understanding_2021, wang_insights_2018} by forming five hydrogen bonds with the dipeptide D-alanyl-D-alanine (D-Ala-D-Ala) in the peptidoglycan layer of the cell wall.\cite{kahne_glycopeptide_2005, wang_insights_2018, salter_infrared_1995, poully_probing_2010, naumann_vibrational_1987} Assmann \textit{et al.}\ reported an antibiotic interaction of vancomycin with Gram-positive \textit{Enterococcus faecalis} after 30~min incubation time using Raman spectroscopy.\cite{assmann_identification_2015} To cover the onset of this interaction, we present sub-cellular, high-resolution chemical contrasts obtained from single-frequency illumination PiF-IR scans of \textit{B.~subtilis} cells harvested after 15, 30, and 60~min incubation with vancomycin. Following the findings of Anindo \textit{et al.}\ on field coupling effects in photothermal imaging of nanostructured surfaces,\cite{anindo_photothermal_2025} we used merges of PiF contrast images from successive scans of the same sample position at two different illumination frequencies for our qualitative discussion including the evaluation of scanning artifacts visible in high-resolution scans and discussing the influence of parameter settings on PiF contrasts. The obtained high-resolution PiF contrasts are compared with recently published results from quick-freeze, deep-etch EM of individual \textit{B.~subtilis} cells.\cite{tulum_peptidoglycan_2019} 

Furthermore, we compare PiF-IR spectra acquired from different positions on the surface of treated and untreated \textit{B.~subtilis} cells harvested after 30~min with FTIR spectra of these samples to demonstrate the surface sensitivity and the spectral sensitivity of PiF-IR. 
We use difference spectra and a chemometrics analysis of the PiF-IR spectra and discuss the results in the context of the known interaction in our model system to evaluate the potential of PiF-IR for studying chemical variations related to antibiotic interaction on the bacteria surface. Additional information on local chemical variations on the surface of treated bacteria cells is obtained by a joint chemometrics analysis of hyperspectral PiF-IR scans in two of the regions presented.

\section{Materials \& Methods}
\subsection{Sample Preparation}
A plain vancomycin sample for IR spectroscopy was prepared from a solution of 1 mg  vancomycin hydrochlorate powder (Sigma-Aldrich) in 1 ml \ce{H2O}. 5 $\mu$l of this aqueous solution were dripped onto a fresh \ce{CaF2} slide and air dried.

\textit{B.~subtilis} subsp.~\textit{spizizenii} (ATCCTM 6633TM) samples were cultivated in CASO broth (ROTH GmbH, Germany) overnight at 37°C while shaking at 160 rpm. The overnight culture was used to inoculate a starter culture (60 ml) with an optical density (OD) of 0.1 at $\lambda = 600$ nm.\cite{assmann_identification_2015} Afterwards, the sample was divided into two 30 ml flasks, shaken for one hour at 30°C and 10 $\mu$g/ml vancomycin were added to one of the flasks. The resulting concentration is the lower range of a typical trough serum concentration recommended for most patients during appropriate vancomycin therapy.\cite{assmann_identification_2015, levine_vancomycin_2006} 1~ml of treated and untreated \textit{B.~subtilis} was harvested at 15, 30 or 60 minutes of each flask. The bacteria samples were then pelleted by centrifugation at 13000 g, washed twice in water, and resuspended in 250 $\mu$l of water. 5 $\mu$l of each sample was carefully placed on a \ce{CaF2} slide and air dried for 30 min at 50°C. The samples were stored at 4°C for up to one week prior to measurement. Pairs of treated and control samples were prepared in three independent experiments and investigated within one week after preparation.

\subsection{FTIR}
FTIR spectra were acquired using a VARIAN 620-IR microscope with a liquid nitrogen-cooled mercury cadmium telluride (MCT) detector at 80 K at a resolution of 2 ${\rm cm}^{-1}$ and in a spectral range of $900 - 4000 {\rm cm}^{-1}$. Firstly, 32 scans of the absorption spectrum of a sample of \textit{B.~subtilis} were acquired at maximum aperture in transmission mode. The presented spectra were then averaged from regions with accumulated bacteria and baseline corrected. The spectra used for the comparison of the methods were vector-normalized (L2 norm).

\subsection{PiF-IR data acquisition \& analysis}
\begin{figure}[t]
\centering
  \includegraphics[width=3.33 in]{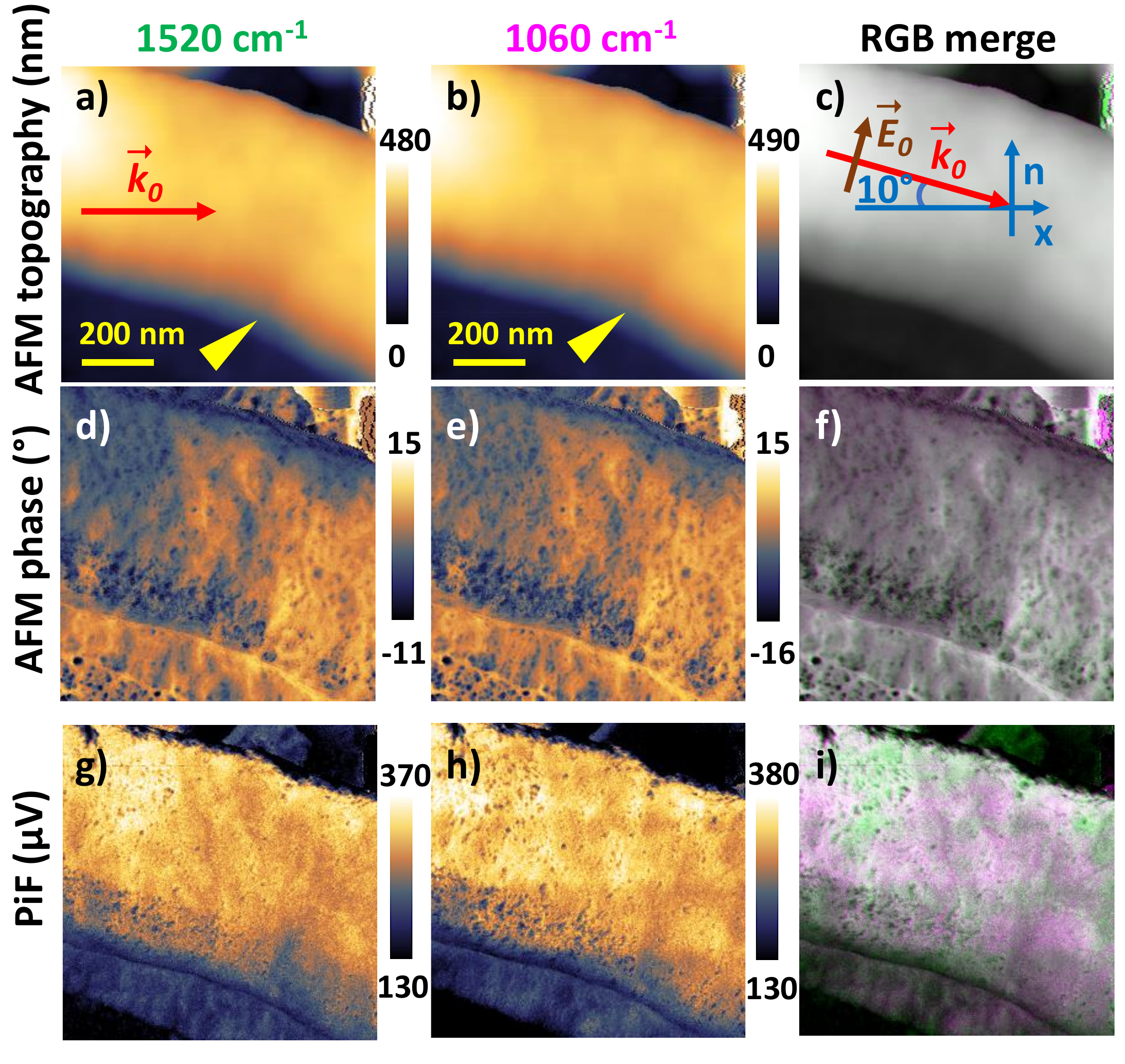}
  \caption{{\bf Single illumination frequency PiF contrasts} of treated \textit{B.~subtilis} harvested after 15~min. a-c) Topography, d-f) AFM Phase and g-i) PiF. Left: $\nu=1520$~${\rm cm}^{-1}$, middle: $\nu=1060$~${\rm cm}^{-1}$, right: RGB with "G" set to $\nu=1520$~${\rm cm}^{-1}$ and "R+B" set to $\nu=1060$~${\rm cm}^{-1}$. Schematics in a indicate the light propagation projected onto the sample plane, and in c the illumination geometry including the electric field oscillation in the plane of incidence normal to the sample plane. The yellow arrows in a,b mark a line that is lower in height than the neighboring cell material.}
  \label{fig1}
\end{figure}
PiF-IR measurements were performed using a VistaScope\textsuperscript{\rm TM} (Molecular Vista Inc., US) equipped with a pulsed quantum cascade laser (QCL) containing 4 QCL chips (Block Engineering, US), which is tunable in the spectral range of 770-1890$~{\rm cm}^{-1}$ at 1~${\rm cm}^{-1}$ spectral bandwidth and illuminates the tip-sample region at a high angle of incidence ($\approx$ 80°).  Ultrasharp, high-frequency, non-contact PtIr coated PointProbeTM Plus (ppp) cantilevers (Nanosensors\textsuperscript{\rm TM}, CH) with a force constant of 42 ${\rm Nm}^{-1}$ and a resonance frequency of 330 kHz were used. The photo-induced force (PiF) was acquired in heterodyne detection mode (sideband), measuring the PiF in the first harmonics of the system $f_1 \approx 300$ kHz, while the cantilever was driven at its second harmonics $f_2\approx$ 1650 kHz. The QCL was pulsed at the frequency $f_m = (f_2 - f_1)$ to enhance the detected signal.\cite{sifat_photo-induced_2022, joseph_nanoscale_2024} The system was operated in non-contact mode by choosing a small amplitude of $\approx 2$ nm for the driven oscillation and a set point of 85\%, which results in probing mainly the attractive regime of the tip-sample interaction force.\cite{jahng_tip-enhanced_2018, jahng_nanoscale_2019} The channel used for driving the oscillation (at $f_2$) provides amplitude feedback and is used to generate topographic (height) images.  

\subsubsection{Compensation of anisotropic intensity distribution in photothermal imaging of nanostructured materials}
Photothermal imaging of nanostructured materials commonly suffers from an anisotropic intensity distribution of the measured signals with respect to the geometry of the nanostructures.\cite{anindo_photothermal_2025, hondl_method_2025} This effect is caused by the geometry of the near-field coupling of the illuminating electric field with the nanostructured materials.\cite{anindo_photothermal_2025} The effect is demonstrated in Fig.~\ref{fig1}, where the two single illumination frequency PiF contrasts obtained in two subsequent scans at $\nu=1520$~${\rm cm}^{-1}$ (Fig.~\ref{fig1}g) and $\nu=1060$~${\rm cm}^{-1}$ (Fig.~\ref{fig1}h) show less intensity on the side of the bacteria cell facing the illumination than on the back side (the propagation vector $k_0$ of the illumination is depicted in the AFM topography Fig.~\ref{fig1}a). An additional inhomogeneity is introduced by the inclination angle of the AFM tip scanning the sample, as is seen in the topography images (Fig.~\ref{fig1}a-c) showing a soft slope in the front and a steep slope in the back of the \textit{B.~subtilis} cell. The yellow arrow marks a line lower than the neighboring cell material, which appears bright in the AFM phase images (Fig.~\ref{fig1}d-f) and dark in the PiF contrasts (Fig.~\ref{fig1}g-i). Consequently, the interpretation of the detected absorption in single-frequency photothermal images is not straightforward and requires a detailed understanding of the illumination geometry in the AFM setup and the related electric field coupling with the nanostructured topography, as well as the impact of the convolution with the shape of the AFM tip\cite{bian_scanning_2021} on the recorded photothermal signal. As demonstrated by Anindo \textit{et al.}, in the case of sufficiently low illumination intensities, these effects do not depend on the choice of the illumination frequency.\cite{anindo_photothermal_2025} Following these findings, we compare the relative PiF intensities obtained in two or three different frequency bands at the same sample position. For this, subsequent scans were cropped using SurfaceWorks\textsuperscript{\rm TM} and used as color channels in RGB merges of PiF contrasts in these bands for our qualitative analysis of local absorption intensities on the surface of \textit{B.~subtilis} cells. As demonstrated in Fig.~\ref{fig1}i, the relative absorption in the two bands and thus the information on the chemical composition is visible independently of the absolute intensity of the recorded PiF.

\subsubsection{Control of mechanical detection frequency in PiF-IR}
The mechanical resonance frequencies of the cantilever shift in the force field of the interaction with a sample with respect to the resonance without any sample.\cite{jahng_tip-enhanced_2018, bian_scanning_2021} Consequently, the resonance frequency will also vary with varying properties of the sample, such as material stiffness, which contribute to the tip-sample interaction force and therefore lead to shifts in $f_1$ in the range of tens of kHz. 
The PIF is evaluated from the measured amplitude at the detection frequency. Thus, a correct evaluation requires the adjustment of the detection frequency. Newer versions of VistaScan\textsuperscript{\rm TM} software provide the possibility of adjusting it to the actual resonance frequency $f_1$ at each data point prior to data acquisition. For further discussion, we refer to the corresponding section on the effect of varying mechanical resonance frequency in the supplementary information.
In our first two scan series, the PiF was recorded at fixed values of the detection frequency, which had been evaluated prior to acquisition at an arbitrary data point. In our series~3, the detection frequency was adjusted at each data point before acquisition. The effect of these settings is discussed in the Results \& Discussion section.

\subsubsection{PiF-IR imaging and spectra acquisition}
PiF contrast images at single illumination frequencies were acquired with an optimized illumination power in the range of a few $100$~$\mu$W resulting in peak intensities between $300 - 600$~$\mu$V, ensuring sufficient contrasts while avoiding substantial heating of the tip, which could result in damage to the sample or the tip coating. The data obtained were inspected for scan artifacts. In some cases, artifacts occurred due to intensity modulations caused by feedback from the instrument’s cooling system during the measurements. Due to the high sensitivity of the method to even small amounts of materials coating the sample or the tip,\cite{jahng_substructure_2018} also contamination during scanning has to be evaluated and controlled. The dynamic non-contact AFM mode used in PiF-IR strongly reduces such contamination. The main cause is insufficient height control during fast scanning. Therefore, using a slow scanning speed even for overview scans is strongly recommended, which, on the other hand, poses a challenge for finding interesting areas in the samples. In general, we carefully examined the artifacts by the eye for mutable impacts on the report of the chemical structure in the scan. Data showing conflicts of scan artifacts with the sample structure were excluded from the analysis except two examples, which are discussed accordingly. The scan data for the treated \textit{B.~subtilis} in series 2 harvested after 15~min (Fig.~\ref{fig2}a) were accepted for analysis after a line-by-line correction of intensity variations in the slow scan direction in SurfaceWorks\textsuperscript{\rm TM}, which improved the quality of the PiF contrast while not interfering with the structure of the sample. Details of this procedure are given together with examples of scan artifacts in the supplementary information. 

For the acquisition of high-resolution scans, we used a slow scan speed of 0.2~lines/s (0.2~$\mu$m/s). Cells of treated \textit{B.~subtilis} were scanned choosing a resolution of 4~nm/pixel, which makes use of the available spatial resolution of PiF-IR in the range of 5~nm,\cite{krishnakumar_nanoscale_2019, murdick_photoinduced_2017, joseph_nanoscale_2024} but resulted in $\approx$1~h per single image (2~h for a single position). The untreated control experiments were scanned using a scan resolution of 8~nm/pixel, which still provides good insight into chemical modifications of the cell surface and reduces the required acquisition time by a factor of 2. The positions of the high-resolutions scans were selected from previously acquired larger area overview scans using a scan resolution of 0.02~$\mu$m/pixel and scanning at a speed of 2~$\mu$m/s.

For each PiF-IR scan, height images (AFM topography) were acquired simultaneously using the instrument feedback at the driving frequency, which matches the second mechanical resonance frequency of the system $f_2$. For the cantilevers used here, the quality factor is typically higher at the first mechanical resonance frequency $f_1$, which is reserved for the detection of the PiF (including also the AFM phase images). As a result, the topography images show less detail than the PiF-IR contrasts and the AFM phase images recorded at $f_1$ (see Fig.~\ref{fig1}) and cannot be compared to AFM topography measurements optimized for good spatial resolution. Furthermore, equipped AFM probes had to match the requirements for sideband detection, involving restrictions in the modulation frequencies of the pulsed QCL, which does not allow the use of instrument parameters tailored for high-resolution AFM topography, such as, for example, the AFM topography of the cell wall of a hydrated \textit{S. aureus} cell presented by Touhami \textit{et al.}\cite{touhami_atomic_2004} All acquired scan images were processed using SurfaceWorks\textsuperscript{\rm TM} 3.0 Release 32 (Molecular Vista Inc, US).

The PiF-IR spectra and hyperspectra were acquired using a homogenized illumination power of about 100 $\mu$W by clipping the previously recorded power spectrum at 5\% of its peak intensity and with a focal spot diameter in the range of twice the illumination wavelength ($2\lambda$). Residual variations in illumination power were removed by calibrating the sample spectra with spectra acquired on non-absorbing \ce{CaF2} substrates. An acquisition time of 70~s was used for single spectra and each pixel in a hyperspectrum. For hyperspectra of 32 x 32 pixels, this accumulates to $\approx20$~h, which is feasible due to the high mechanical stability of the microscope. The scan resolution for the hyperspectra was set to 13~nm/pixel as a compromise between a suitable acquisition time and a still high spatial resolution while covering a spectral range of 250~${\rm cm}^{-1}$ at 1~${\rm cm}^{-1}$ resolution. Due to stability restrictions of the illuminating QCL light source,\cite{joseph_nanoscale_2024} we had to restrict the spectral range of the hyperspectra to that available in one QCL chip only. The presented PiF-IR spectra, hyperspectra, and spectral components were smoothed by a Savitzky-Golay filter 2-11-11. PiF-IR spectra used for method comparison were vector normalized (L2 norm) after removing constant background noise. A chemometric data analysis of PiF-IR spectra and hyperspectra was performed using and extending our homebuilt software \href{https://github.com/BioPOLIM/hyPIRana/}{hyPiRana}.\cite{joseph_nanoscale_2024,maryam_hypirana_2025}

\section{Results \& Discussion}
\subsection{Visualization of fibrillar organization on the surface of cells}
\begin{figure*}[t]
 \centering
 \includegraphics[width=440pt]{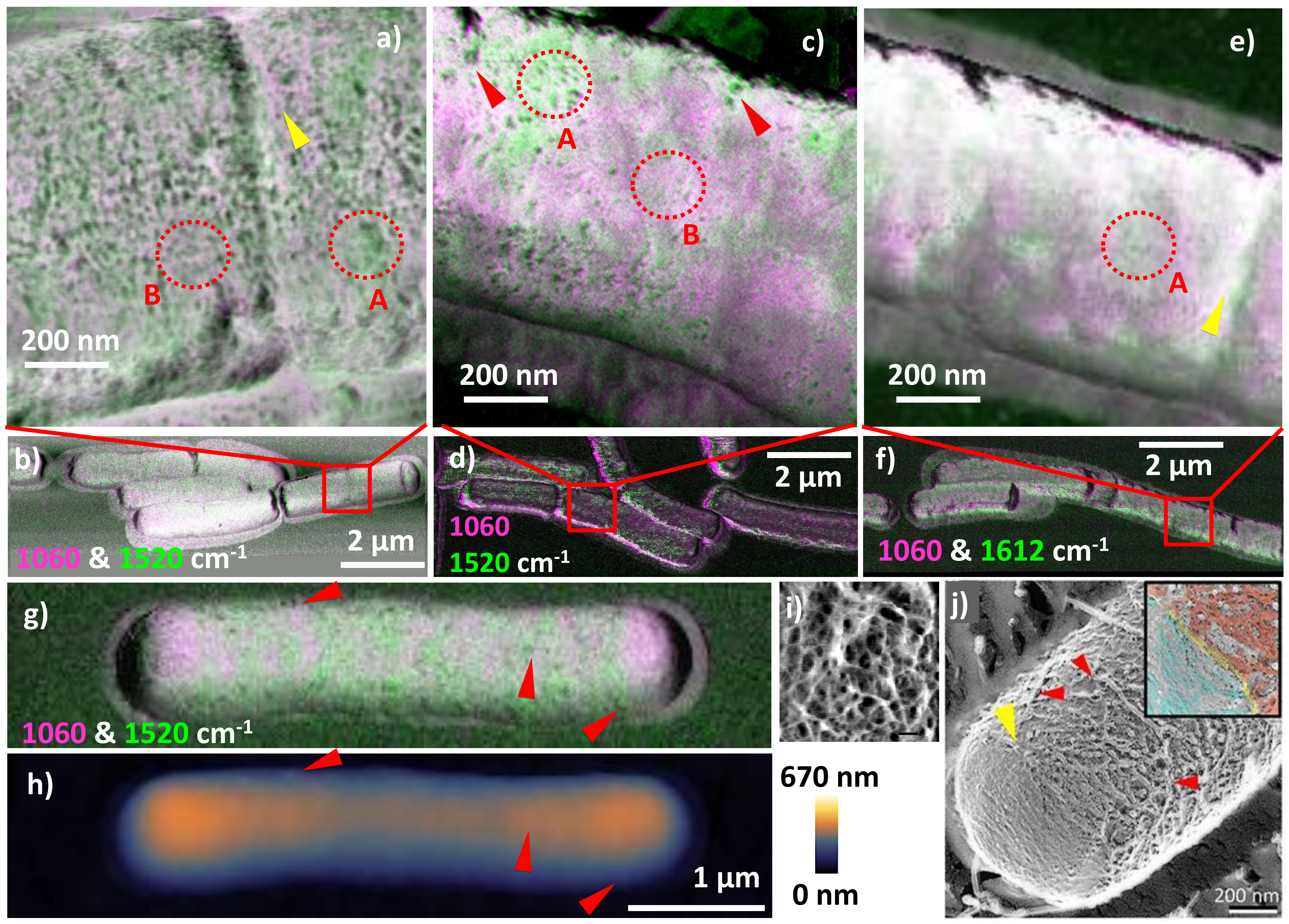}
 \caption{{\bf High-resolution chemical imaging of treated and untreated \textit{B.~subtilis} using PiF-IR.}  a-g) Merged subsequent PiF-IR scan images showing contrasts in glycan (pink) @1060 ${\rm cm}^{-1}$ and peptide (green) @1520 ${\rm cm}^{-1}$ or @1612 ${\rm cm}^{-1}$ as indicated; a-f) \textit{B.~subtilis} incubated with vancomycin for 15~min; yellow arrows in a,e mark piecrusts forming at septa; red arrows in c mark possibly developing septa or depressions; red dotted areas A and B mark losely and densly organized fibrils, respectively. Scans presented in a,c,e, were acquired with increased pixel resolution in the marked areas in the respective overview scans b,d,f using the same IR frequencies. The PiF in a,b,g was recorded at fixed $f_1$, while in c-f $f_1$ it was adjusted in each pixel of the scan. a-d) treated and e,f) untreated \textit{B.~subtilis} (control), g) \textit{B.~subtilis} incubated with vancomycin for 60~min and h) simultaneously acquired topography (at $f_2$), red arrows mark protrusions. Scan resolutions: overviews (b,d,f,g): 0.02 $\mu$m/pixel, zoomed areas: a,c) 4 nm/pixel, e) 8 nm/pixel. i) High-resolution AFM topography image of the cell wall of a hydrated \textit{S.\ aureus} cell (bar: 50 nm); image taken from Touhami \textit{et al.}, Copyright 2004, American Society for Microbiology.\cite{touhami_atomic_2004} j) Surface structures of \textit{B.~subtilis} 168 CA cell visualized by quick-freeze, deep-etch EM: red arrows mark filaments, the yellow arrow marks the boundary between cylindrical and pole parts (bar: 200 nm). Inset: area at the yellow arrow showing cylindrical part (red) pole (blue), and piecrust (yellow); image taken from Tulum \textit{et al.}, \ Copyright 2019, Oxford University Press.\cite{tulum_peptidoglycan_2019}}
 \label{fig2}
\end{figure*}
In the high-resolution PiF-IR contrasts of single \textit{B.~subtilis} cells the organization of peptiodoglycan strands on the cell surfaces becomes visible; see Figs.~\ref{fig2}a and c, which show PiF contrasts of \textit{B.~subtilis} cells obtained as RGB merges from two subsequent PiF-IR scans at scan resolutions of 4 nm/pixel at $\nu=1060$ ${\rm cm}^{-1}$ (pink) and $\nu=1520$ ${\rm cm}^{-1}$ (green). The observations of these irregular structures on the surface of the bacteria agree with the observations of irregular fibrillar structures on the surface of Gram-positive bacteria obtained applying high-resolution AFM to \textit{S.~aureus} by Touhami \textit{et al.}\cite{touhami_atomic_2004} (Fig.~\ref{fig2}i) and quick-freeze deep-etch EM to \textit{B.~subtilis} by Tulum \textit{et al.}\cite{tulum_peptidoglycan_2019} (Fig.~\ref{fig2}j). 
\clearpage
According to Tulum \textit{et al.}, the thickness of individual filaments in the peptidoglycan layer ranges from 6 to 9~nm\cite{tulum_peptidoglycan_2019}, which lies within the spatial resolution of $\approx 5$~nm of PiF-IR.\cite{shcherbakov_photo-induced_2025}
When the scan resolution is reduced by a factor of 2, the contrast in the surface structure is reduced, but not completely lost; see Fig.~\ref{fig2}e.
The most pronounced structural contrast is visible in the image of the bacteria cell belonging to series 2 (Fig.~\ref{fig2}a), which was recorded at a fixed detection frequency resulting in a contribution of opto-mechanic properties to the PiF contrast (see the methods section). The other cells (Figs.~\ref{fig2}c and e) were recorded using a pixel-by-pixel adjustment of the detection frequency. In this case, the chemical variation on the surface of the bacteria cell dominates the contrast, which is the desired information obtained from the investigation using PiF-IR and which we will discuss in the following sections.

\subsection{Subcellular chemical contrasts of \textit{B.~subtilis} cell surface}
The formation of the five hydrogen bonds between vancomycin and the \ce{N-acyl-D-Ala4-D-Ala5} termini of peptidoglycan takes place in a local chemical environment which contains several peptides and other macromolecular structures, which poses a challenge for the discrimination of the bonds. In a first step, we therefore focus on general chemical variations that appear in the surface layer of the \textit{B.~subtilis} cells by comparing PiF contrasts related to glycan absorption with those related to absorption in the amide I and II bands. For this, we used RGB merge images of PiF intensities acquired in two subsequent scans of the same position on the sample obtained at an absorption frequency of glycans ($\nu_1 = 1060$~${\rm cm}^{-1}$, pink, combining red \& blue channels) and at an absorption frequency of amides (either $\nu_2 = 1520$~${\rm cm}^{-1}$ or $\nu_2 = 1612$~${\rm cm}^{-1}$) presented as green channel. In the resulting high-resolution PiF contrasts of two treated \textit{B.~subtilis} cells (Figs.~\ref{fig2}a,c) and an untreated cell (Fig.~\ref{fig2}e) harvested after 15 min, the fibrous surfaces of the peptidoglycan layer are visible, as can be seen by comparing these PiF contrasts with the surface structure of a \textit{B.~subtilis} cell visualized using quick-freeze, deep-etch EM,\cite{tulum_peptidoglycan_2019} which for convenience was reproduced in (Fig.~\ref{fig2}j). The positions of these scans had been selected from overview scans containing several untreated bacteria (Figs.~\ref{fig2}b,d,f). The high PiF intensities appearing at the edge of bacteria cells in two of the overview scans (Figs.~\ref{fig2}d,f) are overshooting artifacts caused by the fast scanning speed.\cite{bian_scanning_2021} Individual bacteria cells were found to be surrounded by flat material that might contain debris from cell walls destroyed during sample preparation that has accumulated around intact cells. As described in the Methods section, the pronounced structural difference in the high-resolution PiF contrasts of the two treated cells (Figs.~\ref{fig2}a,c) is related to the different method settings during data acquisition. 

Cells presented in Figs.~\ref{fig2}c,d (treated) and e,f (control) were harvested from the same experiment after 15~min. Their high-resolution PiF contrasts show local variations in relative intensity in the glycan and amide bands on a scale of 50-100~nm (Figs.~\ref{fig2}c,e). Similar scale variations also appear on top of the fibrous structure in the treated cell of series 2 (Fig.~\ref{fig2}a). This seems to point to a general variation in the surface chemistry of the peptiodoglycan layer on a scale of 50-100~nm. Possible causes could be effects from drying during sample preparation or slight chemical variations during cell growth. The glycan strands appear to be more loosely arranged in some areas showing a higher PiF in the amide bands (red dotted areas A), while more densely organized glycan strands are visible in some areas absorbing in the glycan band (red dotted areas B). However, the available data sets in our current PiF-IR study do not allow for a clear distinction whether this is a general observation. This would require an optimization of the method parameters together with a quantitative analysis of a larger set of samples. The surface structure of the \textit{B.~subtilis} cell investigated using EM reveals an irregular structure of thicker glycan filaments (red arrows in Fig.~\ref{fig2}j) with larger gaps between, while other areas show smaller filaments that are more densely organized, which agrees with the interpretation that variations in the density of the glycan filaments are a general observation of \textit{B.~subtilis} cells and not specific for the antibiotic interaction with vancomycin. In two of the cells, a progressing septa can be seen (Figs.~\ref{fig2}a,e). The piecrusts (yellow arrows) that form at the edge of the septa show a higher intensity in the glycan band ($\nu_1 = 1060$~${\rm cm}^{-1}$, pink) than in the amide bands ($\nu_2 = 1520$~${\rm cm}^{-1}$ or $\nu_2 = 1612$~${\rm cm}^{-1}$, green). This agrees with the observation of thicker glycan strands in piecrusts reported in studies using other methods.\cite{tulum_peptidoglycan_2019} In the septa, we find a generally lower intensity of the glycan band, apart from some fibrous structures extending into the septum from the piecrust. This points to a higher protein content in the reactive material of the progressing septum, which in transition electron micrographs was found to be distinct from the usual cell wall fabric.\cite{touhami_atomic_2004} Some 10-20~nm wide features appearing in the amide II band in the treated cell from series 3 (red arrows in Fig.~\ref{fig2}c) might show the onset of septal formation or cell wall damage; however, the structures are still too small to be clearly discriminated. In general, the peptidoglycan layer seems to be rather intact in untreated and treated cells harvested after 15~min. 
\begin{figure}[t]
\centering
  \includegraphics[width=3.33in]{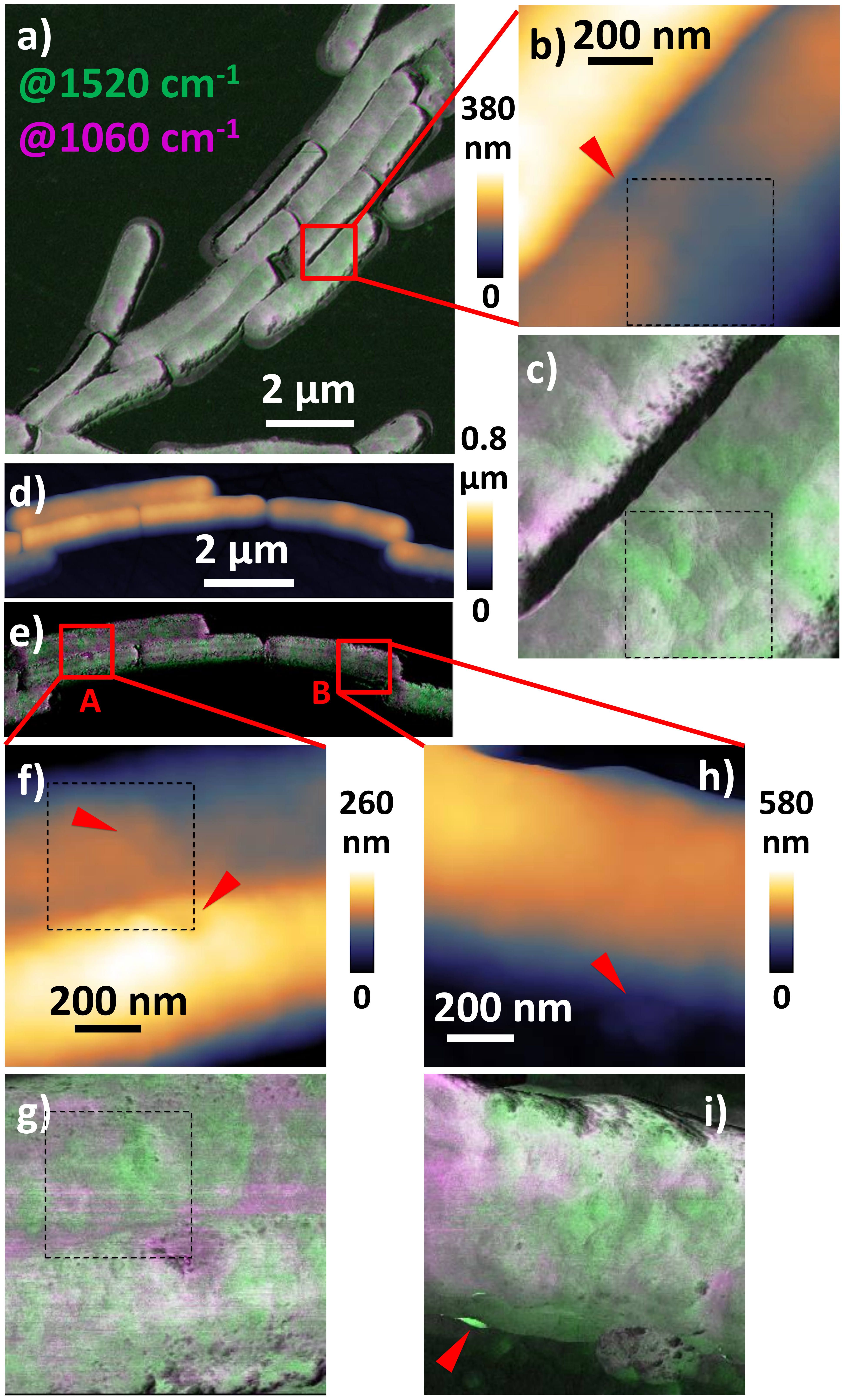}
  \caption{{\bf PiF contrasts of treated \textit{B.~subtilis} harvested after 30 and 60~min from the same experiment.}   a) Overview PiF contrast of \textit{B.~subtilis} cells incubated with vancomycin for 30~min with b) AFM height image in area marked by red square in a) and c) corresponding high-resolution PiF contrast;  d-i) \textit{B.~subtilis} cells incubated with vancomycin for 60~min: d) overview AFM height image, e) corresponding PiF contrast, f,g) AFM height images in areas A and B marked by red squares in e) and h,i) corresponding high-resolution PiF contrasts. Merged subsequent PiF-IR scan images (a,c,e,g,i) show contrasts in glycan (pink) @1060 ${\rm cm}^{-1}$ and peptide (green) @1520 ${\rm cm}^{-1}$ absorption bands. Red arrows in AFM height images mark interesting areas. Red arrow in i marks an artifact. Dashed black boxes mark positions of hyperscans.}
  \label{fig3}
\end{figure}

As expected from the study by Assmann \textit{et al.},\cite{assmann_identification_2015} \textit{B.~subtilis} cells harvested after longer incubation with vancomycin show alterations in the surface of the peptidoglycan layer, which can be clearly associated with cell wall damage. A very pronounced example is seen in the \textit{B.~subtilis} cell incubated with vancomycin for 60~min, presented in Figs.~\ref{fig2}g and h. In this particular cell, several protrusions visible in the simultaneously acquired topography image Fig.~\ref{fig2}f correspond to areas showing a higher PiF intensity in the amide II band than in the glycan band, examples are marked by red arrows. Their sizes are in the range of 100-200~nm, which could point to the formation and release of \textit{B.~subtilis} EVs.\cite{brown_extracellular_2014, kim_extracellular_2016, toyofuku_prophage-triggered_2017} \textit{B.~subtilis} EVs are known to form from the bacteria membrane underneath the peptidoglycan layer, which is consistent with the observation of an enhanced PiF intensity at 1520~${\rm cm}^{-1}$ (green) of these protrusions in the PiF contrast in Fig.~\ref{fig2}e. 
\clearpage

Various alterations in the cell surface, which can be associated with cell wall damage, are already visible in the PiF contrasts of several treated \textit{B.~subtilis} cells harvested after 30~min incubation with vancomycin presented in Fig.~\ref{fig3}a (this image is reproduced in Fig.~\ref{fig4}a together with the corresponding AFM height image Fig.~\ref{fig4}b, which includes markups for positions of acquisitions of PiF-IR spectra). We conducted high-resolution scans on a cell showing pronounced variations in PiF contrast in the position marked by a red square in the overview image in Fig.~\ref{fig3}a. The AFM height image (Fig.~\ref{fig3}b) of this marked position shows a depression on the surface of this cell that indicates damage to the cell wall (red arrow). In the corresponding area, the PiF contrast image (Fig.~\ref{fig3}c) shows a higher intensity in the amide II band (1520~${\rm cm}^{-1}$, green) than in the glycan band (1060~${\rm cm}^{-1}$, pink). This observation could be related to the inhibition of peptidoglycan synthesis by the presence of vancomycin in the position of a starting cell division, which prevents the formation of a piecrust and exposes the underlying cell membrane. However, the PiF contrast of this cell reveals several areas with high absorption in the amide II band, which are not all indications of a progressing cell division but would also agree with an interpretation as an early stage of cell wall destruction leading to the release of EVs. 

Our investigation of \textit{B.~subtilis} harvested after 60~min confirmed several features of progressing cell death. An example showing protrusions in one \textit{B.~subtilis} cell is presented in Fig.\ref{fig2}g,h, as discussed above. In another experiment, we performed high-resolution PiF-IR scans in two areas (A and B) that showed variations in cell height in the AFM height overview image (Fig.~\ref{fig3}d), see the red squares marking the positions in the corresponding PiF contrast image (Fig.~\ref{fig3}e). A depression line in position A (red arrow in the upper part of Fig.~\ref{fig3}f) corresponds to an area in the PiF contrast (Fig.~\ref{fig3}g) showing a higher absorption in the amide II band (520~${\rm cm}^{-1}$, green) than in the glycan band (1060~${\rm cm}^{-1}$, pink). In contrast, the round depression seen on the surface of the neighboring cell predominantly absorbs in the glycan band (pink). Unfortunately, an artifact appears as horizontal lines that absorb predominantly in the glycan band in the PiF contrast images of these positions. Particularly strong lines appear in the middle of Fig.~\ref{fig3}g. Their equidistant spacing in the slow scan direction (vertical in the view) can be explained by material moved by the AFM tip over the cell surface during the previous faster overview scans of these positions. The lines become weaker in the lower part of the scan after scanning the round depression. Thus, a great deal of the moved material seems to have accumulated in that depression, probably also causing the highly absorbing spot at its center. Within our qualitative analysis of cell wall features from several treated cells, we did not find a similar structure showing high absorption in the glycan band. The observed depressions in the cell wall appear to be correlated rather with a higher absorption in the amide bands, which underlines the caution required to interpret this singular observation. The cell in position B is thicker (higher) in the left part of the scan (Fig.~\ref{fig3}h), which corresponds to higher absorption in the glycan band compared to the remaining cell surface (Fig.~\ref{fig3}i) and agrees with a higher protein content in regions where the underlying cell membrane is exposed due to damage to the cell wall. The protrusion in the lower part of the scan (red arrow in Fig.~\ref{fig3}h) predominately absorbs in the amide II band. Dark spots on its surface correspond to height modulations. At the steep edge of the bacteria cell, the appearance of convolution effects of the AFM tip with the sample topography, as well as artifacts due to instabilities in tip oscillations, have to be considered. The latter is likely the cause of the two spots showing a very high intensity in the amide II band in the lower left of the cell (red arrow). The above findings of variations in the cell wall of \textit{B.~subtilis} will be used to guide the evaluation of spectral variations obtained by chemometrics of PiF-IR spectra of cells incubated with vancomycin for 30~min in the following section.

\subsection{Chemometrics of PiF-IR spectra from treated and untreated \textit{B.~subtilis} cells harvested after 30 min}
\begin{figure*}[t]
 \centering
 \includegraphics[width=400pt]{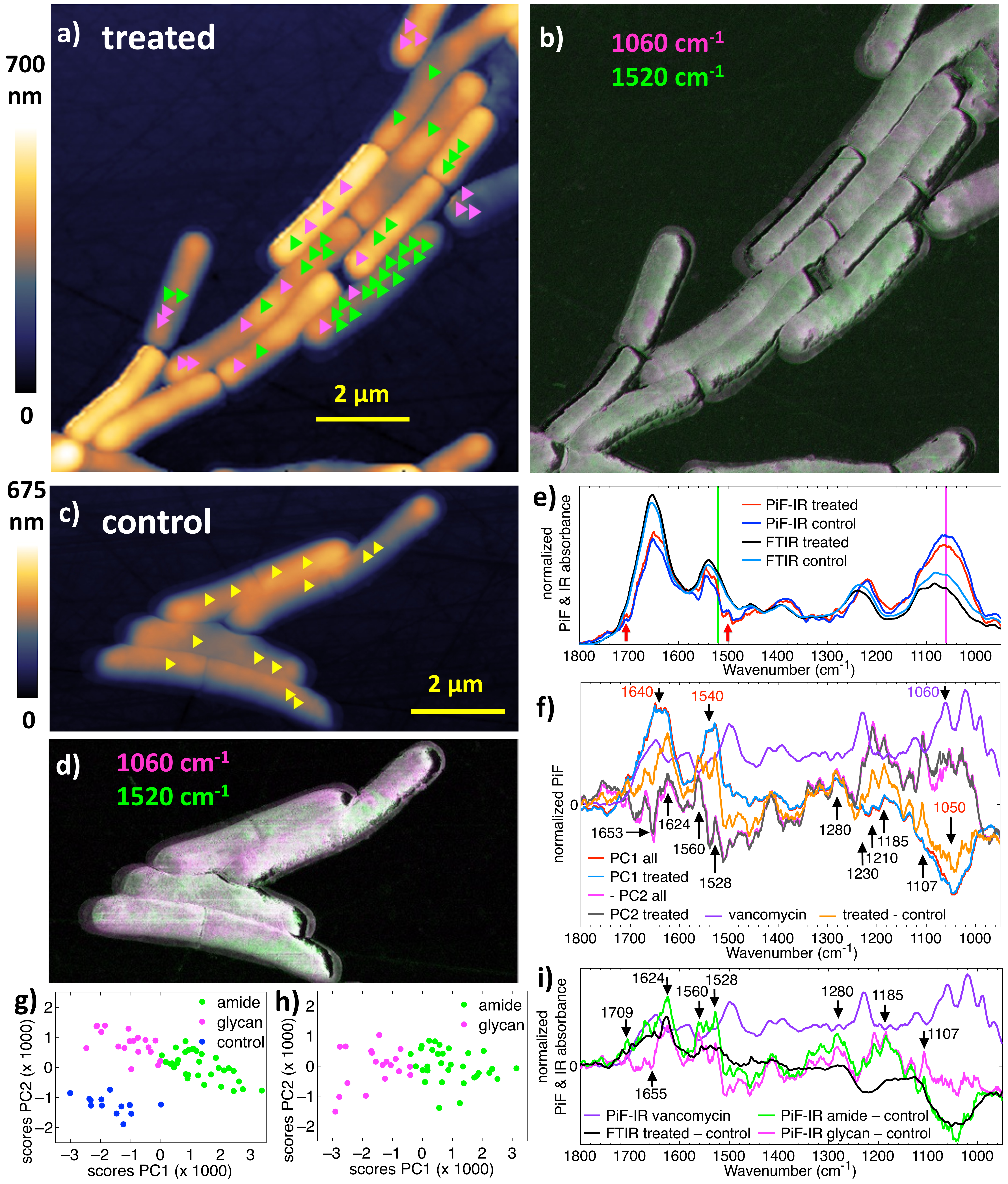}
 \caption{{\bf PiF-IR spectra and chemometrics of \textit{B.~subtilis} cells harvested after 30 min: a-d) Scan images:} AFM topography of a) treated cells and c) control sample, positions of point spectra acquisition are marked by triangles following the color code of the “glycan” and “amide” subgroups found by chemometrics; b,d) PiF contrasts from pairs of subsequent sans acquired @1520 ${\rm cm}^{-1}$ (green) and @1060 ${\rm cm}^{-1}$ (pink); e) mean spectra of PiF-IR spectra acquired on treated (a,b) and control sample (c,d) and corresponding FTIR spectra, red arrows mark shoulder peaks in PiF-IR spectra; {\bf f-i) difference spectra and chemometrics:} f) 1st and 2nd loadings of PCA applied to all PiF-IR spectra and of PCA applied to spectra from treated cells only, PiF-IR spectrum of vancomycin and difference spectrum of PiF-IR spectra of treated and control presented in e; g) and h) scores maps of PC1 and PC2 obtained from the PCA on all PiF-IR spectra and on spectra of treated cells, respectively. Spectra of treated cells were sorted in “glycan” (pink) and “amide” (green) according to positive and negative scores in the PC1 from analysis of all PiF-IR spectra; i) difference spectra of subsets.}
 \label{fig4}
\end{figure*}

We acquired PiF-IR spectra in the indicated positions in two sample areas containing several cells harvested after 30~min from the same experiment (Figs.~\ref{fig4}a-d). Both samples show variations in the PiF contrasts at the selected frequencies, see Fig.~\ref{fig4}b for the treated \textit{B.~subtilis} and Fig.~\ref{fig4}d for the untreated control. Focusing on variations in treated cells, we acquired 51 and 13 individual PiF-IR spectra on manually selected positions on treated and untreated cells, respectively, which are marked by triangles in the simultaneously acquired AFM topography images Figs.~\ref{fig4}a and c. The color code for the spectra of treated cells follows the assignment into the “amide” (green) and “glycan” (pink) subsets resulting from our chemometric analysis of all PiF-IR spectra, see the discussion below. 

Unfortunately, the PiF contrasts of the control were affected by artifacts resulting from tip contamination during scanning; for details, see Fig.~S2 and the related discussion in the supporting information. PiF-IR spectra were acquired in less contaminated areas. Our selection is approved by comparison of normalized mean PiF-IR spectra with normalized FTIR spectra obtained from the same \textit{B.~subtilis} strains (Fig.~\ref{fig4}e), which confirms the expected difference upon treatment with vancomycin: In the spectra obtained in both methods, a slightly increased intensity is seen for the treated \textit{B.~subtilis} cells in respect to the untreated cells (Fig.~\ref{fig4}e) in the amide I band at about 1650~${\rm cm}^{-1}$ and the amide II band at about 1540~${\rm cm}^{-1}$ as well as a decrease in the broad absorption band at about 1060~${\rm cm}^{-1}$ corresponding to carbohydrates and phosphodiesters. This finding agrees with a reduced contribution from peptidoglycan as a result of damage to the cell wall in treated cells. From this we conclude that important spectral variations caused by vancomycin treatment can nonetheless be obtained from the analysis of these PiF-IR spectra in spite of the contamination appearing in the scans of the control sample.
\clearpage
The spectral resolution of the average PiF-IR spectra is higher than that of the FTIR spectra. For example, the red arrows in Fig.~\ref{fig4}e mark shoulder peaks in the amide I and II regions, which are not resolved in the broad peaks in the FTIR spectra. Due to the high spatial resolution of PiF-IR, the probed sample volume is several orders of magnitude smaller than that of FTIR, resulting in less averaging of chemically heterogeneous samples.\cite{shcherbakov_photo-induced_2025} Additionally, the band positions in the PiF-IR spectra are slightly red-shifted to frequencies smaller than those of the FTIR spectra. Such shifts are commonly observed as a result of differing illumination geometries in infrared spectroscopy methods. In the instruments used here, the variation is between the normal incidence in FTIR and an $\approx 80\degree$ oblique incidence in PiF-IR. Details on the influence of this effect require geometric modeling of the electromagnetic near-field including the plasmonic enhancement of the metallic tip used in PiF-IR, which so far has been investigated only in general,\cite{anindo_photothermal_2025, jahng_quantitative_2022, sifat_photo-induced_2022} but not for such spectroscopic details. 

A further, quite prominent feature in this comparison of normalized spectra is the comparably enhanced intensity in the amide bands of the FTIR spectra compared to that of the PiF-IR spectra, which can be explained by the differing probe volumes in the methods. Because of its high spatial sensitivity, the PiF-IR signal is restricted to the surface of the bacteria, which is dominated by carbohydrates and phosphodiesters of the peptidoglycan layer. FTIR probes complete bacteria cells including the lipoproteins in the cell membrane and proteins in the cytoplasm\cite{davies-jones_photoinduced_2023} resulting in a higher concentration of amides in the probed volume compared to that probed in PiF-IR. Additional effects contributing to the differences between the FTIR and PiF-IR spectra may be caused by the high sensitivity of tip-enhanced methods to molecular alignment\cite{joseph_nanoscale_2024, quaroni_mid-infrared_2018, ruggeri_infrared_2021} and band orientation.\cite{anindo_photothermal_2025, bakir_orientation_2020, luo_intrinsic_2022}

In the next step, we will present the evaluation of spectral variations in these data sets using the difference spectra of the spectra of treated and untreated \textit{B.~subtilis} discussed above and complement them by an unguided chemometrics of the PiF-IR spectra data sets. The results are discussed in the context of literature reports on the spectral signature of the known interaction. In the difference spectrum of the average PiF-IR spectra (yellow line in Fig.~\ref{fig4}f), a strong peak appears at 1624~${\rm cm}^{-1}$ in the amide I spectral region, which is also visible in the FTIR difference spectrum (black line in Fig.~\ref{fig4}i) on top of a broader peak. According to Barth \textit{et al.}, hydrogen bonding lowers the frequency of the amide I peak by $\approx 20$~${\rm cm}^{-1}$\cite{barth_infrared_2007} which supports the assignment of this band to the known interaction of vancomycin with peptidoglycan. In addition to this strong peak, the PiF-IR difference spectrum shows a fine structure, which is hardly visible in the FTIR difference spectrum and results from highly localized sampling in PiF-IR\cite{shcherbakov_photo-induced_2025}. These spectral details were reproduced in our chemometrics analysis and will be discussed in the following including both approaches. 

We conducted two principal component analyses (PCAs), one on all PiF-IR spectra and the other one on those of treated cells only; for the scores see Fig.~\ref{fig4}g and h, respectively. In both data sets, the loading of the first component (PC1, red and light blue lines in Fig.~\ref{fig4}f) shows a pronounced anticorrelated variation of the amide bands and the broad absorption band at 1050~${\rm cm}^{-1}$, see Fig.~\ref{fig4}f, which matches the variation reported by the PiF-IR and FTIR difference spectra; see the yellow line in Fig.~\ref{fig4}f and the black line in Fig.~\ref{fig4}i, respectively. The control sample shows negative PC1 scores (Fig.~\ref{fig4}g) in accordance with the expected higher content of glycan components on the surface of untreated cells. In contrast, the cluster of spectra from treated cells covers a wide range of negative and positive PC1 scores. This finding agrees with the pronounced local chemical variations of the surfaces of individual bacteria cells in the corresponding PiF contrast image (Fig.~\ref{fig4}b). Several cells show considerable damage to their peptidoglycan layer exposing the proteinaceous underlying cell membrane, resulting in a higher PiF intensity in the amide II band than in the glycan spectral region. From this we conclude that the first PC presents the chemical difference between the peptidoglycan layer and the underlying cell membrane, and thus reveals the extent of cell wall damage in the positions of the corresponding spectra in the sample. Following this observation, we divided the treated spectra into an “amide” group (green dots) and a “glycan” subgroup (pink dots in Fig.~\ref{fig4}g), exhibiting, positive and negative PC1 scores, respectively. The cluster split according to this assignment was reproduced in the PCA of treated cells only, confirming the dominant effect of cell wall damage on the chemical variation in these spectra. A comparison of the corresponding color code of the marked positions of the acquired spectra in the topography image (Fig.~\ref{fig4}a) with the PiF contrast in those positions (Fig.~\ref{fig4}b) shows good agreement of most assignments with the PiF contrast in the two illumination frequencies. The variation of PC1 scores in the control sample (blue dots in Fig.~\ref{fig4}g) may be explained by a general chemical heterogeneity of the peptidoglycan layer surface as discussed with the samples harvested after 15~min incubation. Additional variations could stem from damage to the cell wall resulting from sample preparation (due to pelleting and drying) and possibly may also contain some contribution of the smeared material during scanning.

Having assigned PC1 to the major chemical variation between damaged and intact cell wall regions, we now will discuss the second pronounced chemical variation in the data set reported by the second principal component (PC2) and in the PiF-IR difference spectra. The clear distinction between treated and untreated cells in PC2 (see the scores map in Fig.~\ref{fig4}g) indicates the assignment of the spectral variation in PC2 to vancomycin interaction. Indeed, the PC2 loadings (Fig.~\ref{fig4}f) show several matching bands, as will be discussed in the following. Vancomycin is known to form five hydrogen bonds with the \ce{N-acyl-D-Ala4-D-Ala5} termini in the peptidoglycan cell wall.\cite{kahne_glycopeptide_2005} Three of these are formed with the carboxylate group of the tripeptide and two between amide groups of both vancomycin and the tripeptide.\cite{kahne_glycopeptide_2005, poully_probing_2010}  The absorption of $\alpha$-helices typically appears around 1650-1665~${\rm cm}^{-1}$ in the amide I band.\cite{joseph_nanoscale_2024,davies-jones_photoinduced_2023,barth_infrared_2007,fabian_infrared_2006} According to Barth \textit{et al.}, the formation of hydrogen bonds causes a red-shift of $\approx 20$~${\rm cm}^{-1}$ of this absorption.\cite{barth_infrared_2007} This agrees with the observation of a sharp dip at 1653~${\rm cm}^{-1}$ next to a broader peak at 1624~${\rm cm}^{-1}$ in the PC2 loadings as well as in the difference spectrum (Fig.~\ref{fig4}f). The formation of hydrogen bonds with $\beta$ sheets also contributes to the peak at 1624~${\rm cm}^{-1}$\cite{barth_infrared_2007}. The comparison of the PiF-IR difference spectra of either the amide or the glycan subgroup and the control (green and pink lines in Fig.~\ref{fig4}f, respectively) confirms the assignment of this new absorption to the formation of hydrogen bonds with the peptidoglycan, as the new band appears not only on top of the amide I absorption in the amide subgroup, but very prominently in the glycan subgroup, in which the broad absorption of glycans at 1050~${\rm cm}^{-1}$ is almost not altered with respect to the control. 

In the amide II band, a blue-shift of about 10~${\rm cm}^{-1}$ is expected upon the formation of hydrogen bonds between the amide groups in vancomycin and \ce{N-acyl-D-Ala4-D-Ala5}.\cite{poully_probing_2010} This agrees with the observation of two sharp peaks at 1560~${\rm cm}^{-1}$ and 1528~${\rm cm}^{-1}$ accompanied by two sharp dips at 1540 and 1510~${\rm cm}^{-1}$ in the PC2 loadings (Fig.~\ref{fig4}f), which again are more prominent in the glycan subgroup (Fig.~\ref{fig4}i). The corresponding alterations are seen as a weak substructure on top of the amide II peak in the FTIR difference spectrum (Fig.~\ref{fig4}i).

Due to the overlap of \ce{C=O} symmetric stretching vibrations with \ce{CH2} bending and \ce{CN} stretching vibrations in the amide III spectral region ($1400-1200$~${\rm cm}^{-1}$), a detailed discussion of the various peaks and dips that appear in this region is not feasible. 
The bands appearing at $\nu>1700$~${\rm cm}^{-1}$ (lipid carbonyl stretch) and at 1107 and $\approx 1220$~${\rm cm}^{-1}$ (\ce{PO2}-stretching) in the PC2 loadings and in the PiF-IR difference spectra might be related to phospholipids in the cytosol membrane,\cite{kochan_vivo_2018} which is exposed upon destruction of the peptidoglycan cell wall. 
We will use the above findings on the vancomycin interaction signature in PiF-IR spectra for our analysis of PiF-IR hyperspectra of two treated \textit{B.~subtilis} cells presented in the following.

\subsection{Localizing vancomycin interaction in hyperspectra of treated \textit{B.~subtilis}}
\begin{figure*}[ht]
 \centering
 \includegraphics[width=460pt]{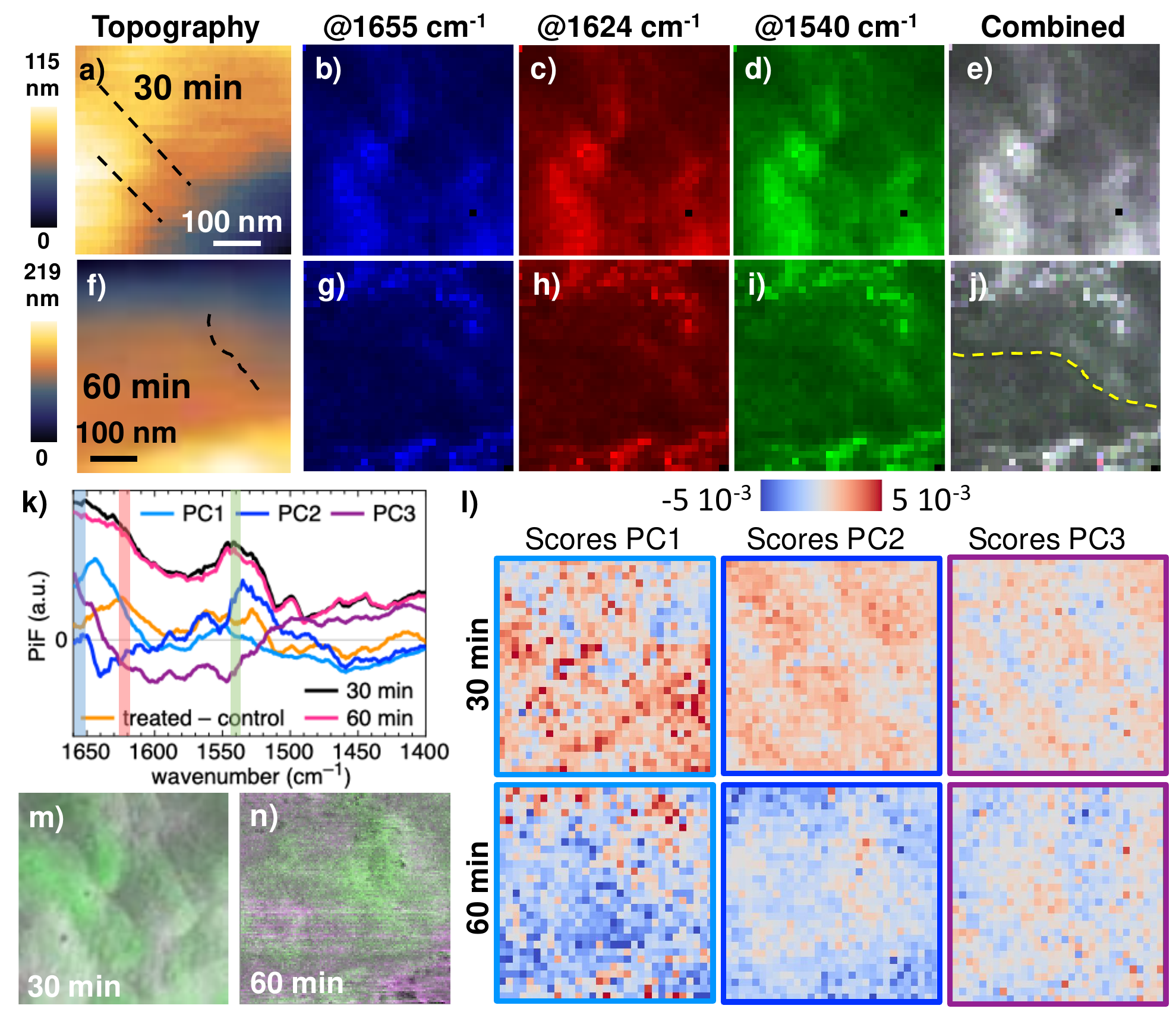}
 \caption{{\bf PiF-IR Hyperspectra of treated \textit{B.~subtilis} harvested after a-e) 30~min and f-j) 60~min from the same experiment:} a,f) AFM topography with red lines indicating the positions of spectra presented in k) and l), respectively, dashed black lines in a) mark depression lines; b-d) and g-i) RGB channels showing PiF intensities at selected frequencies: $1655\pm2$~${\rm cm}^{-1}$ (blue), $1624\pm2$~${\rm cm}^{-1}$ (red), and $1540\pm2$~${\rm cm}^{-1}$ (green); the bands are marked in corresponding colors in the plotted spectra in k); e,j) combined RGB images; artifacts from scanning appeared below the yellow line in j). k,l) combined PCA of the two PiF-IR hyperspectra: k) PC loadings together with mean spectra and the difference spectrum from single spectra (Fig.~\ref{fig4}f), and l) PC scores maps; m,n) high-resolution PiF contrasts of the positions of the hyperpectra cut from the dashed boxes in Figs.~\ref{fig3}c and g, respectively.}
 \label{fig5}
\end{figure*}
To localize the vancomycin interaction on the surface of treated \textit{B.~subtilis} cells, we acquired hyperscans covering the spectral range of $1400-1660$~${\rm cm}^{-1}$ in positions showing cell wall destruction in high-resolution PiF contrast images of \textit{B.~subtilis} cells treated for 30~min and for 60~min (marked by dashed black boxes in Figs.~\ref{fig3} c and g, respectively). In accordance with the single spectra chemometric results, we selected the band at 1624~${\rm cm}^{-1}$ to visualize the formation of hydrogen bonds between vancomycin and \ce{N-acyl-D-Ala4-D-Ala5}; see PC2 loadings and PiF-IR difference spectra in Figs.~\ref{fig4}f,i. As mentioned above, there is an influence of the nanostructured topography on the PiF intensity,\cite{anindo_photothermal_2025} and the untreated \textit{B.~subtilis} cells also show some absorption at 1624~${\rm cm}^{-1}$, as can be seen in the FTIR and mean PiF-IR spectra of treated and control samples in Fig.~\ref{fig4}e. Thus, it is not feasible to discern locations of vancomycin interaction from positions showing an enhanced PiF signal at 1624~${\rm cm}^{-1}$ alone. However, we can locate vancomycin interaction positions using PiF contrasts with two other bands at 1655~${\rm cm}^{-1}$ and 1540~${\rm cm}^{-1}$, which present amide absorption in the absence of hydrogen bonding with vancomycin resulting in local dips in the difference spectra in Figs.~\ref{fig4}f,i. Therefore, we selected three spectral bands at $1655\pm2$~${\rm cm}^{-1}$, $1624\pm2$~${\rm cm}^{-1}$, and $1540\pm2$~${\rm cm}^{-1}$ in each hyperspectrum and set them as the blue, red and green channels of an RGB image, respectively; see Figs.~\ref{fig5} b-d and g-i. 

The three channels of each position show rather similar contrasts, resulting in overall grayscale RGB images, with structures showing slightly higher absorption in each of the channels appearing in both combined images (Figs.~\ref{fig5}e,j). As expected, the spatial variation of the PiF intensity in the three channels and the combined image is related to structures in the topography. For example, depression lines marked by dashed black lines in the AFM topography images (Figs.~\ref{fig5}a and f; for higher resolutions scans of these positions, see Figs.~\ref{fig3}b and f) appear dark in all three PiF contrasts and the combined images (Figs.~\ref{fig5}b-e, and g-j). Because of the choice of the bands, variations in the overall content of amides will show up simultaneously in all three channels. These variations are also reported by the PC1 of the combined PCA of the two hyperscans: The low scores seen in blue in the scores maps (left column in Fig.~\ref{fig5}l) appear dark in the PiF contrasts of all channels and in the combined images. The PC1 loading contains two broad bands at $\approx 1550$~${\rm cm}^{-1}$ and $\approx 1640$~${\rm cm}^{-1}$ (light blue line in Fig.~\ref{fig5}k) which overlap with the bands in the PC1 of the single spectra (Fig.~\ref{fig4}f), and which cover the spectral ranges of the three selected channels (marked by vertical lines in corresponding colors in Fig.~\ref{fig5}k. In the PiF contrasts of the cell treated for 60~min (Figs.~\ref{fig5}g-j), a dark area appears in the bottom left, which correlates with low scores in the corresponding PC1 scores map (Fig.~\ref{fig5}l, bottom left). This area is part of the position, where pronounced contamination from scanning has been observed (marked by the dashed yellow line in the combined RGB image Fig.~\ref{fig5}j.) As discussed above, in this area scan lines absorbing at 1060~${\rm cm}^{-1}$ appeared in the PiF contrasts in Figs.~\ref{fig3}e and g, which result from material smeared by the tip over the sample during scanning. For convenience, a cut of the high-resolution PiF contrast (dashed black box in Fig.~\ref{fig3}g), which matches the area of the hyperscan is presented in Fig.~\ref{fig5}n. The lower absorption of amides in this region of the quite coarse hyperscan ($32\times32$~pix) is likely to be related to contamination of the surface of the bacteria cell by some material absorbing in the glycan band (1060~${\rm cm}^{-1}$) and therefore this area will not be considered for further discussion.

We expect to localize vancomycin interaction in positions showing an enhanced intensity at 1624~${\rm cm}^{-1}$ (red channels, Figs.~\ref{fig5}c,h) and thus appearing in reddish color in the combined RGB images (Figs.~\ref{fig5}e,j). In the PiF-IR hyperscan of the \textit{B.~subtilis} harvested after 30~min incubation with vancomycin, an increased intensity in the red channel (1624~${\rm cm}^{-1}$) is visible in the combined image (Fig.~\ref{fig5}e) above and below the two depression lines marked by dashed black lines in the simultaneously acquired topography image (Fig.~\ref{fig5}a) and also in the high-resolution PiF contrast of this position; see Fig.~\ref{fig3}c and the cut of the corresponding area presented in Fig.~\ref{fig5}m. As discussed above, this area might show a distorted septum that suffers from inhibited peptidoglycan synthesis caused by the interaction with vancomycin. The formation of a piecrust at the edges of the septum has been stopped by the hydrogen bonds between vancomycin and \ce{N-acyl-D-Ala4-D-Ala5}, which agrees with the observed enhanced absorption at 1624~${\rm cm}^{-1}$ in this region.
\clearpage

The area between the two depression lines shows an increased intensity in the green channel (1540~${\rm cm}^{-1}$), which is mainly related to the \ce{N-H} bending vibrations of the amide groups\cite{filip_ft-ir_2004,barth_infrared_2007,fabian_infrared_2006} in the absence of hydrogen bond formation between the amide groups in vancomycin and \ce{N-acyl-D-Ala4-D-Ala5} which would cause a blue-shift of about 10~${\rm cm}^{-1}$.\cite{poully_probing_2010} This area also shows an enhanced intensity at 1520~${\rm cm}^{-1}$ in respect to that at 1060~${\rm cm}^{-1}$ in the high-resolution PiF contrast of this position (Fig.~\ref{fig5}m). In treated and untreated \textit{B.~subtilis} cells harvested after 15~min, we had observed an enhanced PiF intensity at 1520~${\rm cm}^{-1}$ in respect to that at 1060~${\rm cm}^{-1}$ inside forming septa (in Figs.~\ref{fig2}a and e), which supports the interpretation that the hyperscan of this position shows the distorting effect of vancomycin on a forming septum. Similarly, the depression line on the surface of the cell harvested after 60~min and the surrounding area show a higher intensity in the green channel (at $1540\pm2$~${\rm cm}^{-1}$) and also an enhanced intensity at 1520~${\rm cm}^{-1}$ in the high-resolution PIF contrast (Fig.~\ref{fig5}n) is seen. There are also pixels with higher absorption in the red channel (at $1624\pm2$~${\rm cm}^{-1}$) in this area and adjacent to it, but the structure appears to be less organized than in the area surrounding the depression lines in the cell harvested after 30~min. This may be related to a higher degree of disintegration of the cell wall of the cell incubated with vancomycin for 60~min.

In addition to the higher absorption in the green channel observed between the two depression lines in the cell treated for 30~min, the lower left and upper right of this hyperscan also show enhanced absorption in this band (Fig.~\ref{fig5}e). All three regions have high scores of PC2 appearing in red color in the corresponding scores map (Fig.~\ref{fig5}l, upper middle). The PC2 loading and the difference spectrum of the treated and control single spectra (dark blue and yellow line in Fig.~\ref{fig5}k, respectively) appear to be anticorrelated in the three selected spectral bands (marked by lines of the corresponding colors). However, their shape agrees in other spectral regions, for example in the peak at $\approx 1560$~${\rm cm}^{-1}$ and in decreased intensities between 1430 and 1470~${\rm cm}^{-1}$. From this we conclude that positive scores in PC2 do not simply mark areas of chemical composition similar to the control sample, but also contain materials that were not present on the surface of untreated \textit{B.~subtilis} cells and that are not affected by hydrogen bonds between vancomycin and \ce{N-acyl-D-Ala4-D-Ala5}. Further chemical variations also appear inside these areas, as can be seen in the score maps of the third component of the PCA (Fig.~\ref{fig5}l, right column), which reveal additional spatial variation compared to the PC2 (Fig.~\ref{fig5}l, middle column). However, a detailed discussion of these chemical variations requires further investigation of the cytosol membrane material underneath the peptidoglycan layer, which is beyond the scope of this work.

\section*{Conclusions}
In this work we demonstrated the power of using contrasts obtained in two different spectral bands for a qualitative analysis of the local chemical composition in photothermal imaging of microbes based on recent findings from modeling and experimentation on nanostructured materials by Anindo \textit{et al.}.\cite{anindo_photothermal_2025} A quantitative evaluation of  photothermal images is still not straightforward and requires further work on standardization including the evaluation of artifacts, which for example result from contamination of the sample position due to tip-sample contact in contact and intermittent AFM-IR, and even in true non-contact AFM-IR in cause of overshooting in fast-scanning.

The application of PiF-IR to the well-known model system \textit{B.~subtilis} and vancomycin for antibiotics acting on cell wall synthesis demonstrates the potential of PiF-IR to complement the high structural information of EM and high-resolution AFM, as well as the chemical information of conventional IR spectroscopy by providing unprecedented spatial information on the chemical composition of microbe surfaces. The qualitative evaluation of PiF contrasts from untreated and treated \textit{B.~subtilis} cells confirms the presence of cell wall destruction in treated cells after 30~min incubation with vancomycin in agreement with the observation reported by Assmann \textit{et al.}\ in their study using Raman spectroscopy.\cite{assmann_identification_2015} The high spatial resolution provided using PiF-IR enables the chemical characterization of particular types of destruction, including observation of the underlying cytosol membrane in damaged regions and also the appearance of protrusions that could present the release of EVs from a \textit{B.~subtilis} cell harvested after 60~min treatment with vancomycin.

The comparison of average PiF-IR spectra with FTIR spectra obtained from treated and untreated \textit{B.~subtilis} cells harvested after 30~min from the same experiment shows agreement in the band positions and in the effect of vancomycin treatment, while demonstrating the higher spectral sensitivity and surface sensitivity of PiF-IR, resulting from the several orders of magnitude smaller probe volume in PiF-IR, which reduces the probed volume to the cell surface. Small shifts of the peak positions between FTIR and PiF-IR spectra can be understood as an effect of the different illumination geometry on band positions, which is familiar from comparisons of IR spectra obtained using FTIR with normal incidence and ATR with oblique incidence of the IR illumination. The chemometrics analysis of the PiF-IR spectra of these cells revealed spectral variations in several bands that can be assigned to the formation of hydrogen bonds between vancomycin and its target \ce{N-acyl-D-Ala4-D-Ala5} in the peptidoglycan layer of the cell wall.

The power of PiF-IR to localize the interaction with vancomycin on the surface of treated \textit{B.~subtilis} cells and its effect on the progress of cell division is demonstrated exemplarily in a cell harvested after 30~min incubation which shows a distorted septum in high-resolution PiF contrasts. Using PiF contrasts in three selected spectral bands in a hyperspectrum of this region, the formation of hydrogen bonds with vancomycin becomes visible in the distorted piecrusts enveloping a pair of depression lines visible in the topography of the cell. The material between the lines shows an absorption similar to that found for the material inside the septa observed in treated and untreated \textit{B.~subtilis} cells harvested after 15~min from two different experiments.

The drawback of the very long acquisition time of almost 1 day in hyperspectral images in PiF-IR might be reduced in further studies of known materials by restricting the spectral range to selective bands only.

\section*{Acknowledgments}
D. Täuber acknowledges funding by a postdoctoral Scholarship (PolIRim) from the Friedrich Schiller University Jena in 2020, and from Profile Line LIGHT, Friedrich Schiller University Jena (pintXsum) in 2022. The authors thank the Leibniz-Institute of Photonic Technology for providing access to the VistaScope instrument, for which financial support of the European Union via the Europ\"{a}ischer Fonds für Regionale Entwicklung (EFRE) and the “Th\"{u}ringer Ministerium für Wirtschaft, Wissenschaft und Digitale Gesellschaft (TMWWDG)” (Project: 2018 FGI 0023) is highly acknowledged.

\section*{Author contributions}
Maryam Ali: Writing, Data Analysis, Programming. Robin Schneider: Experimental, Data Analysis, Writing. Nila Krishnakumar: Experimental. Anika Strecker: Experimental, Data Analysis. Sebastian Unger: Programming. Mohammad Soltaninzehad: Programming. Johanna Kirchhoff: Experimental. Astrid Tannert: Experimental. Katerina A. Dragounova: Experimental. Rainer Heintzmann: Methodology, Programming, Editing. Anne-Dorothea Müller: Experimental. Christoph Kraft: Experimental. Ute Neugebauer: Conceptualization, Methodology, Writing. Daniela Täuber: Conceptualization, Writing, Methodology, Experimental, Data Analysis.

\section*{Conflicts of interest}
There are no conflicts to declare.

\section*{Supporting Information}
A short summary of the theoretical framework in photo-induced force microscopy based on the considerations by [Jahng \textit{et al., Phys. Rev. B}, 2022, \textbf{106}, 155424] and [Anindo \textit{et al., J. Phys. Chem C}, 2025, \textbf{129}, 4517] followed by a discussion of the effect of varying the mechanical detection frequency and examples of scan artifacts.

\section*{Data availability}
Data for this article, including data sets obtained using PiF-IR and FTIR are available at Zenodo at: https://doi.org/10.5281/zenodo.14959278.\cite{schneider_mid-infrared_2025} The code used for the chemometrics of PiF-IR hyperspectra in this work can be found at Zenodo at https://doi.org/10.5281/zenodo.15270457.\cite{maryam_hypirana_2025} The code used for this study is also available as release v2.0.0 of our homebuilt software code \textit{hyPIRana} on github: https://github.com/BioPOLIM/hyPIRana.
\clearpage

\section{SUPPORTING INFORMATION\\ Nano-chemical cell-surface evaluation in photothermal spectroscopic imaging of antimicrobial interaction in model system \textit{Bacillus subtilis} \& vancomycin}

\subsection*{Supporting Information: abstract}
The Supporting Information provides a short summary of the theoretical framework in photo-induced force microscopy based on the considerations by [Jahng \textit{et al., Phys. Rev. B}, 2022, \textbf{106}, 155424] and [Anindo \textit{et al., J. Phys. Chem C}, 2025, \textbf{129}, 4517] followed by a discussion of the effect of varying the mechanical detection frequency and examples of scan artifacts.

\tableofcontents


\addcontentsline{toc}{section}{PiF-IR signal detection in side band mode}
\subsection{PiF-IR signal detection in side band mode}
In the following, we will provide a short summary of the photoinduced force ($F_\text{pi}$) in photoinduced force microscopy (PiFM, including PiF-IR) operated in heterodyne side-band mode. PiFM evaluates the gradient of the distance-dependent interaction force $F_{\text{ts}}$ between the AFM tip and the sample. A general expression for $F_{\text{ts}}$ is given by~\cite{ sifat_photo-induced_2022, jahng_quantitative_2022, anindo_photothermal_2025}
\begin{equation}
    F_{\text{ts}}(z)=F_\text{c}(z)+F_{\text{nc}}(z)
\end{equation}
where $F_\text{c}(z)$ describes the contribution from conservative and $F_{\text{nc}}(z)$ non-conservative forces and $z$ is the direction normal to the sample plane. The conservative forces are given by
\begin{equation*}
    F_\text{c}(z) \approx 
    \begin{cases}
        \frac{-H_{\text{eff}}r}{12z^2}  & \text{for} \ (z>r_0) \\
        \frac{-H_{\text{eff}}r}{12 r_0^2}+\frac{4}{3}E^*\sqrt{(r_0-z)^3 r} \ \ \ \ &\text{for} \ (z< r_0)
    \end{cases}
\end{equation*}
$H_{\text{eff}}$ is the effective Hamaker constant which is a measure for the strength of the interaction energy between two bodies via a medium, $r$ is the tip apex radius and $E^*$ is the effective Young's or elastic modulus which is a measure of the stiffness and rigidity of the material. When the distance between the tip and the ample is less than the interatomic distance $r_0$, the conservative force becomes the sum of two forces. The first term is the non-contact van der Waals force and the second term is the contact Derjaguin, Muller, and Toporov (DMT) force.~\cite{jahng_quantitative_2022}.

In the heterodyne detection mode, $F_\text{pi}$ is detected in non-contact AFM mode, i.e. employing a high setpoint of the cantilever oscillation and a small oscillation amplitude $A_2\approx 1-2$~nm for the oscillation driven at the second mechanical resonance frequency $f_2$ of the cantilever. $F_\text{pi}$ is obtained at the first mechanical resonance frequency $f_1$, and the illuminating laser is modulated at $f_{m} = f_{2}-f_{1}$.\cite{jahng_quantitative_2022, anindo_photothermal_2025} In the small oscillation limit, $F_\text{pi}$ can be given as~\cite{jahng_nanoscale_2019, anindo_photothermal_2025}
\begin{equation}\label{eq.thermal force}
    F_\text{pi} \approx \frac{\delta F_{\text{ts}}}{\delta z}\Delta z,
\end{equation}
where $F_{\text{ts}}$ is the distance-dependent interaction force between the AFM tip and the sample and $\Delta z$ is the photoinduced thermal expansion normal to the sample plane.

\addcontentsline{toc}{section}{Effect of varying mechanical resonance frequency}
\subsection{Effect of varying mechanical resonance frequency}
\begin{figure}[h]
\centering
  \includegraphics[height=13 cm]{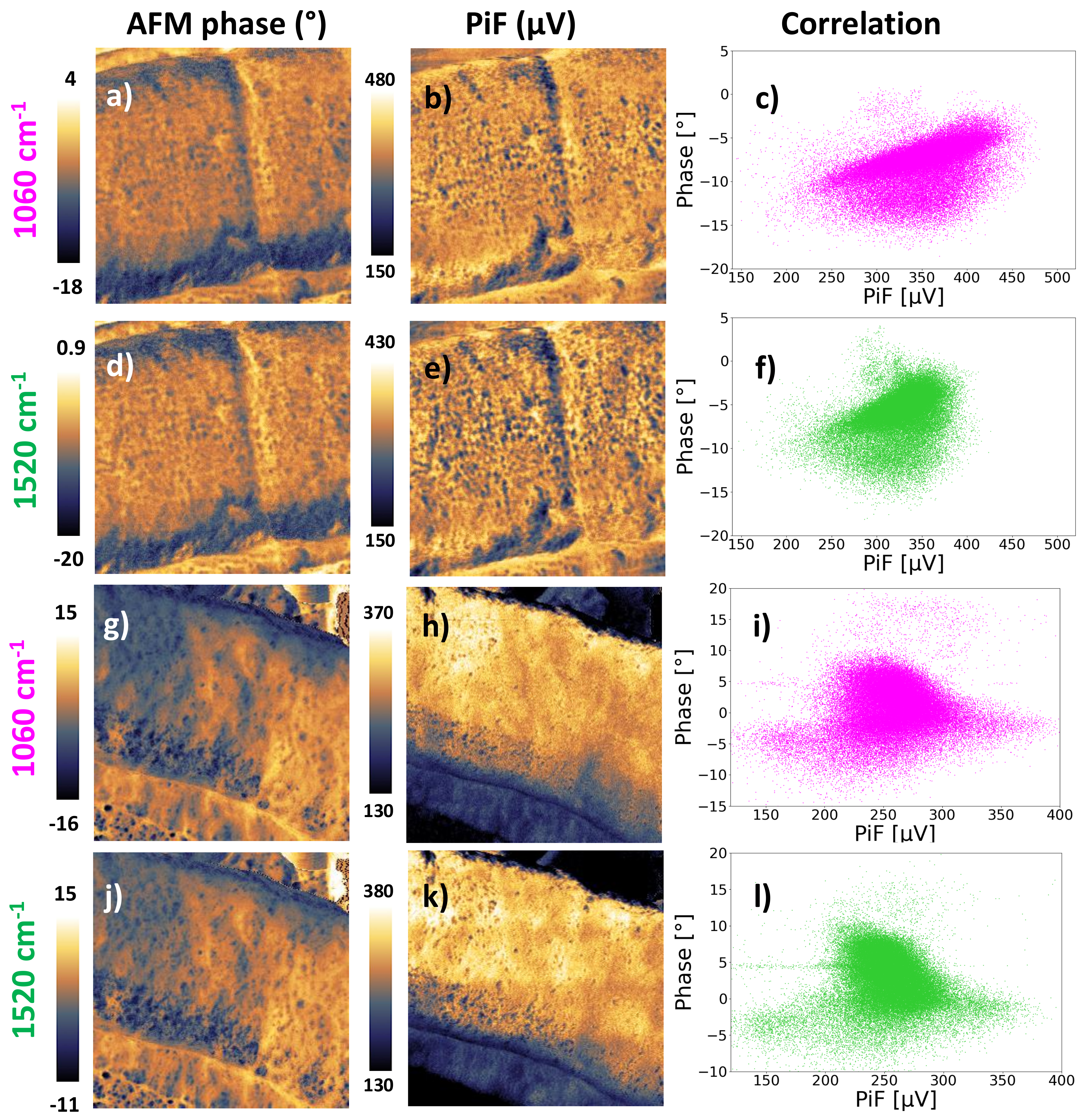}
  \caption{{\bf Correlations between AFM phase and PiF contrasts of treated \textit{B.\ subtilis} harvested after 15 min.} a-c) and d-f) scan of cell of series 2 acquired @1060 ${\rm cm}^{-1}$ and @1520 ${\rm cm}^{-1}$, respectively. g-i) and j-l) scan of cell of series 3 acquired @1060 ${\rm cm}^{-1}$ and @1520 ${\rm cm}^{-1}$, respectively. AFM phase images: a,d,g,j, PiF contrasts: b,e,h,k and their pixel-wise correlations: c,f,i,l.}
  \label{fig.S1}
\end{figure}
It is well known that mechanical properties such as stiffness and viscosity influence the phase as well as the resonance frequencies in dynamic AFM measurements.~\cite{bian_scanning_2021} A change in these properties can be recognized by a corresponding phase shift between the driving oscillation and the resulting oscillation of the cantilever. Such shifts are presented in the AFM phase-contrast images of the scanned sample areas.

In our first two series of PiF-IR measurements, the mechanical resonance frequency of the system was evaluated once prior to scanning a sample area for PiF signal acquisition. As a result, the strength of the recorded PiF signal is modulated by the distance to the actual resonance frequency at the particular position in the scan. As the AFM phase shifts together with the resonance frequency, this can be seen by comparing the acquired PiF contrasts with the AFM phase contrasts in the same position. An example is given in Figs.~\ref{fig.S1}a-f showing AFM phase and PiF contrasts of two subsequently acquired high-resolution scans of a treated \textit{B. subtilis} cell harvested a 15~min from our series 2 together with pixel-wise correlations of PiF and AFM phase. For both illumination frequencies, a predominately linear correlation is seen between PiF intensities and the simultaneously acquired phase phase of the cantilever oscillation. In contrast, a linear correlation is not observed in the corresponding data sets of our series 3, which were acquired with an adjustment of $f_1$ in each pixel prior to acquisition (Figs.~\ref{fig.S1}g-l).

\addcontentsline{toc}{section}{Examples of scan artifacts in PiF-IR}
\subsection{Examples of scan artifacts in PiF-IR}
\begin{figure}[htbp]
\centering
  \includegraphics[height=9 cm]{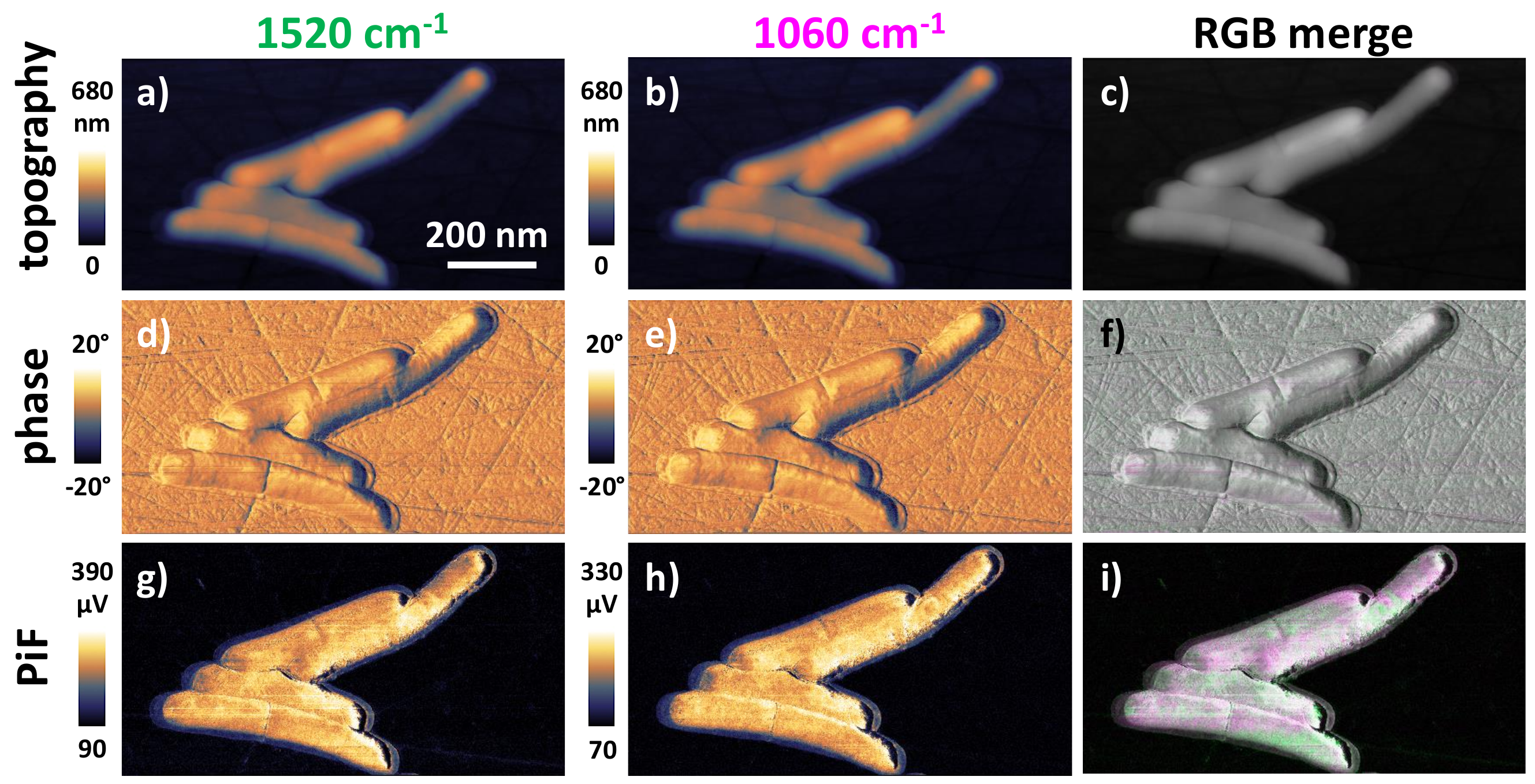}
  \caption{{\bf Subsequently acquired PiF contrasts of untreated \textit{B.~subtilis} harvested after 30 min, series 3 together with RGB merges}. a-c) AFM topography, d-f) AFM phase and g-j) PiF acquired @1520 ${\rm cm}^{-1}$ (a,d,g) and @1060 ${\rm cm}^{-1}$ (b,e,h).}
  \label{fig.S2}
\end{figure}
In AFM several types of artifacts are known to appear.\cite{ricci_recognizing_2004, ukraintsev_artifacts_2012, canale_recognizing_2011} An example is sample material attached to the AFM probe during scanning. This typically results in either a recurring pattern appearing in the image in the case of a particle adhered to the tip, or in lines in the fast scanning direction in the case of loose materials temporarily attached to the tip. The latter was observed in subsequent scans of the untreated \textit{B. substilis} cells harvested after 30~min from our series 3, see Fig.~\ref{fig.S2}. Horizontal lines appear in the AFM phase and in the PiF contrasts of both scans but not in the lower-resolution topography images (acquired at the second mechanical resonance frequency). As discussed in the methods section, overshooting of the tip oscillation at the steep edge of bacteria cells may cause tip-sample contact even in dynamic non-contact AFM. Obviously, this occurred during the first scan at 1520 ${\rm cm}^{-1}$ resulting in sharp horizontal lines in the AFM phase and in the PiF contrast of the scan; see Figs.~\ref{fig.S2}d and g, respectively. Several of these lines also appear in the subsequent scan at 1620 ${\rm cm}^{-1}$ (Figs.~\ref{fig.S2}e and h), which shows that the material was not only temporarily attached to the tip, but also smeared over the surface of the sample during scanning. Due to the high surface sensitivity of PiF-IR even small amounts of material smeared over the surface show pronounced effects on the signal intensity, as can be seen in the RGB merge (Fig.~\ref{fig.S2}i) of the two subsequently acquired PiF contrasts. However, the lines of the smeared material are narrow and therefore could be avoided for the acquisition of PiF-IR spectra on that sample.

\begin{figure}[h]
\centering
  \includegraphics[height=12 cm]{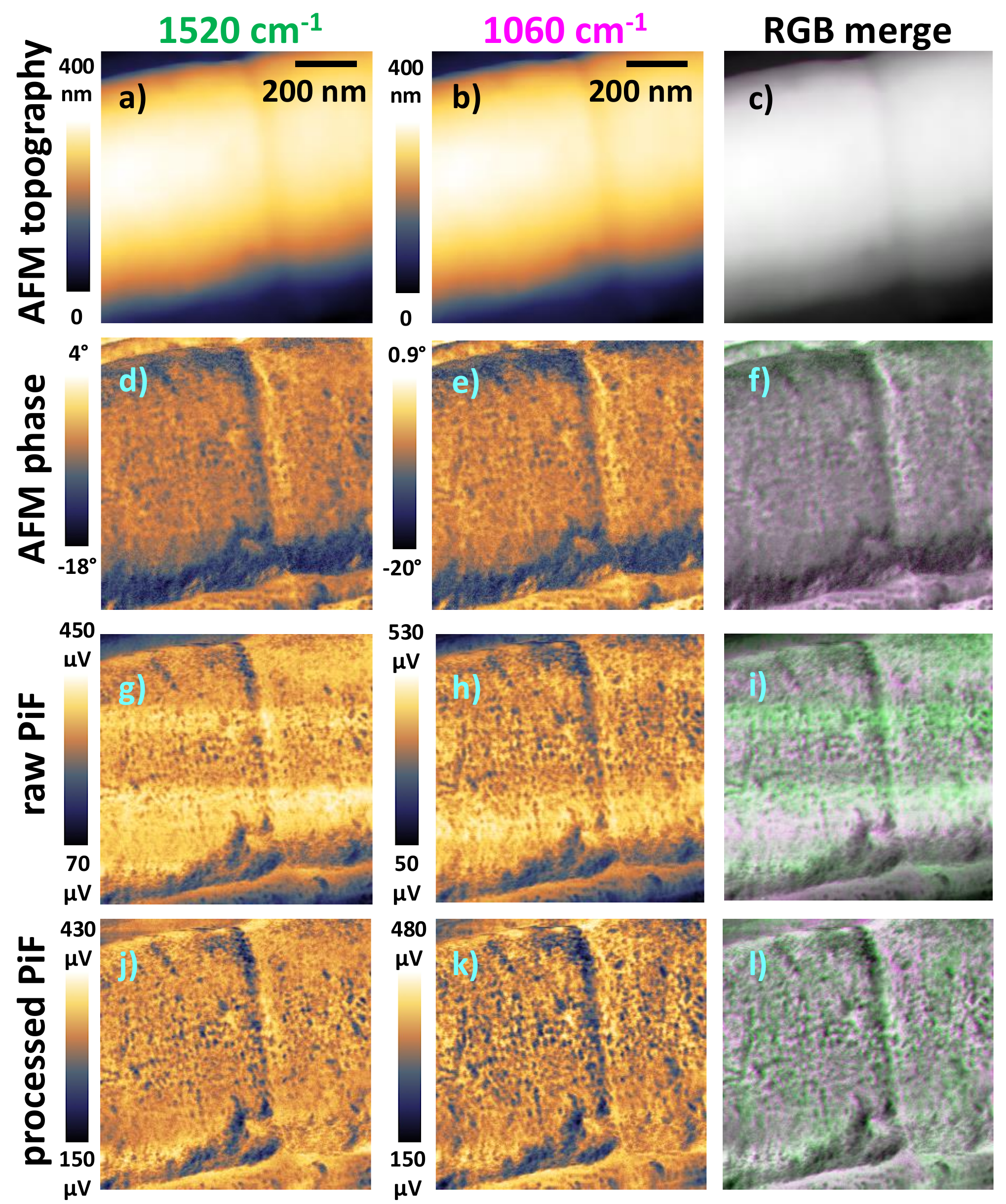}
  \caption{{\bf PiF signal instability in subsequently acquired PiF contrasts of treated \textit{B.~subtilis} harvested after 15 min, series 2, together with RGB merges}. a-c) AFM topography, d-f) AFM phase, g-j) acquired PiF and j-l) line wise corrected PiF acquired @1520 ${\rm cm}^{-1}$ (a,d,g,j) and @1060 ${\rm cm}^{-1}$ (b,e,h,k).}
  \label{fig.S3}
  \end{figure}
In PiF-IR scanning artifacts may also be introduced by instabilities in the alignment of the laser beam illuminating the tip-sample region of by power instabilities of the light source. In some cases, we had observed feedback from the cooling system on the PiF intensity in our instrument. An example is presented in Fig.~\ref{fig.S3} showing subsequent scans of a treated \textit{B.~subtilis} harvested after 15 min from our series 2. The periodic instability in signal intensity resulted in a pattern of equidistant horizontal stripes in single PiF contrasts (Figs.~\ref{fig.S3}). It is quite unlikely that the next scan will start in the same phase of the periodic signal. Therefore, the pattern differs in subsequent scans. However, such a broad horizontal stripe pattern can be easily corrected for by a line-wise correction using the mean intensity of each line, a feature frequently used to improve scanning probe microscopy images. Figs.~\ref{fig.S3}j, k and i show the PiF contrasts after the application of this correction for the subsequent scans and the RGB merge, respectively. The corrected images improve the chemical contrast, which was already visible in the raw images. In this particular sample position, the PiF contrast appears to be predominantly perpendicular to the fast scanning direction, and therefore is not affected by this kind of image processing. In case of a chemical contrast parallel to the fast scanning direction, a similar image processing could cause a conflict with the signal of interest.

\bibliography{PiFIR-Bsubtilis} 

\end{document}